# Can phones, syllables, and words emerge as side-products of cross-situational audiovisual learning? — A computational investigation

Khazar Khorrami
Unit of Computing Sciences, Tampere University, Finland

Okko Räsänen
Unit of Computing Sciences, Tampere University, Finland
Department of Signal Processing and Acoustics, Aalto University, Finland

**Abstract:** Decades of research has studied how language learning infants learn to discriminate speech sounds, segment words, and associate words with their meanings. While gradual development of such capabilities is unquestionable, the exact nature of these skills and the underlying mental representations yet remains unclear. In parallel, computational studies have shown that basic comprehension of speech can be achieved by statistical learning between speech and concurrent referentially ambiguous visual input. These models can operate without prior linguistic knowledge such as representations of linguistic units, and without learning mechanisms specifically targeted at such units. This has raised the question of to what extent knowledge of linguistic units, such as phone(me)s, syllables, and words, could actually emerge as latent representations supporting the translation between speech and representations in other modalities, and without the units being proximal learning targets for the learner. In this study, we formulate this idea as the so-called *latent language hypothesis* (LLH), connecting linguistic representation learning to general predictive processing within and across sensory modalities. We review the extent that the audiovisual aspect of LLH is supported by the existing computational studies. We then explore LLH further in extensive learning simulations with different neural network models for audiovisual cross-situational learning, and comparing learning from both synthetic and real speech data. We investigate whether the latent representations learned by the networks reflect phonetic, syllabic, or lexical structure of input speech by utilizing an array of complementary evaluation metrics related to linguistic selectivity and temporal characteristics of the representations. As a result, we find that representations associated with phonetic, syllabic, and lexical units of speech indeed emerge from the audiovisual learning process. The finding is also robust against variations in model architecture or characteristics of model training and testing data. The results suggest that cross-modal and cross-situational learning may, in principle, assist in early language development much beyond just enabling association of acoustic word forms to their referential meanings.

**Corresponding author(s):** Okko Räsänen, Unit of Computing Sciences, Tampere University, P.O. Box 553, FI-33101, Tampere, Finland. Email: okko.rasanen@tuni.fi.

**ORCID ID(s):** https://orcid.org/0000-0002-0537-0946

## Introduction

When learning to communicate in their native language, infants face a number of challenges that they need to overcome in order to become proficient users of the language. In order to understand speech, they need to figure out how to extract words from the running acoustic signal and how the words relate to objects and events in the external world (cf. Quine, 1960). In order to develop syntactic skills and become creative and efficient users of the language, they must understand that speech is made of units smaller than individual words, allowing combination of these units to form new meanings. In essence, this means that the child learner has to acquire understanding of spoken language as a hierarchical compositional system. In this system, smaller units such as phonemes or syllables make up larger units such as words and phrases, and where these units are robust against different sources of non-phonological variability in the acoustic speech.

The journey from a newborn infant without prior linguistic knowledge to a proficient language user consists of several learning challenges. While one body of developmental research has investigated how infants can utilize distributional cues related to phonetic categories of their native language (e.g., Werker and Tees, 1984; Maye et al., 2002; see also Kuhl et al., 2007 for an overview), another set of studies has focused on the question of how infants could segment acoustic word forms from running speech where there are no universal cues to word boundaries (e.g., Cutler and Norris, 1988; Mattys et al.,1999; Saffran et al., 1996; Thiessen et al., 2005; Choi et al., 2020). Yet another line of research has investigated word meaning acquisition, assuming that words as perceptual units are already accessible to the learner. In that research, the focus has been on the details of the mechanisms that link auditory words to their visual referents when they co-occur at above-chance probability across multiple infant-caregiver interaction scenarios (e.g., Smith and Yu, 2008; Trueswell et al., 2013; Yurovsky et al., 2013), also known as cross-situational learning.

All these different stages have received a great deal of attention in the existing research, both experimental and computational. However, we still have limited understanding on how the different stages and sub-processes in language learning interact with each other, what drives learning in all these different tasks, and what type of acoustic or linguistic representations infants actually develop at different stages of the developmental timeline. For instance, does adaptation to phonetic categories pave the way for lexical development (cf. NLM-e framework by Kuhl et al., 2017), or is early lexical learning a gateway to refined phonemic information (cf. PRIMIR theory by Werker & Curtin, 2005)? How accurately do words have to be segmented before their referential meanings can be acquired?

In contrast to viewing language learning as a composition of different learning tasks, an alternative picture of the process can also be painted: what if processes such as word segmentation or phonetic category acquisition are not necessary stepping stones for speech comprehension, but that language learning could be bootstrapped by meaning-driven predictive learning, where the learner attempts to connect the (initially unsegmented) auditory stream to the objects and events in the observable surroundings (Johnson et al., 2010; Räsänen and Rasilo, 2015; also referred to as discriminative learning in Baayen et al., 2015; see also Ramscar and Port, 2016). While tackling this idea has been challenging in empirical terms, a number of computational studies have explored this idea along the years (e.g., but not limited to, Yu et al., 2005; Roy and Pentland, 2002; Räsänen and Rasilo, 2015; Chrupała et al., 2017; Alishahi et al., 2017; Räsänen and Khorrami, 2019; ten Bosch et al., 2008; Ballard and Yu, 2004). These models have demonstrated successful learning of speech comprehension skills in terms of connecting words in continuous speech to their visual referents with minimal or fully absent prior linguistic knowledge.

Since rudimentary semantics of spoken language seem to be accessible to (computational) learners without having to first learn units such as phone(me)s, syllables or words, it is of interest whether some type of representations for such units could actually *emerge as a side-product* of the cross-modal and cross-situational learning process. The idea is that, instead of learners separately and sequentially tackling a number of sub-problems on the road towards language proficiency, linguistic knowledge could emerge as a latent representational system that effectively mediates the "translation" between auditory speech and other internal representations related to the external world or the learner itself. While not precluding the fact that certain aspects of language skills are likely to emerge earlier than others, the key value of this idea—here referred to as latent language hypothesis (LLH)—is that it replaces a number of proximal language learning goals (phoneme category learning, word segmentation, meaning acquisition) with a unified learning goal of minimizing the predictive uncertainty in the multisensory environment of the learner. This goal aligns well with the popular view of the mammalian brain as a powerful multimodal prediction machine (Friston, 2010; Clark, 2013; see also Meyer and Damasio, 2009, or Bar, 2011), and also fits to the picture of predictive processing at various levels of language comprehension (e.g., Warren, 1970; Jurafsky, 1996; Jurafsky et al., 2001; Watson et al., 2008; Kakouros et al., 2018; Cole et al., 2010). Even if cross-modal learning would not be the primary mechanism for acquisition of linguistic knowledge, it is important to understand the extent the cross-modal dependencies can facilitate (or otherwise affect) the process.

The goal of this paper is to review and explore the feasibility of LLH as a potential mechanism for bootstrapping the learning of language representations at various levels of granularity without ever explicitly attempting to learn such representations. We specifically focus on the case of audiovisual associative learning between visual scenes and auditory speech. We build on the existing computational studies on the topic, and attempt to provide a systematic investigation of LLH by comparing a number of artificial neural network (ANN) architectures for audiovisual learning. We first define LLH in terms of high-level computational principles and review the existing research on the topic in order to characterize the central findings so far. We then present our computational modeling experiments of visually-grounded language learning, where we investigate a large battery of phenomena using a unified set of evaluation protocols: the potential emergence of phone(me)s, syllables, words, and word semantics inside the audiovisual networks. We study whether individual artificial neurons and layers of neurons become correlated with different linguistic units, and whether this leads to qualitatively discrete or continuous nature of acquired representations in terms of time and representational space. Finally, we summarize and discuss our findings and the extent that the LLH could explain early language learning.

While our experiments largely rely on existing body of work in this area (see section Earlier Related Work), our current contributions include i) a coherent theoretical framing of the present and earlier studies under the concept of LLH, ii) an integrative summary of the existing research, iii) systematic experiments investigating several different aspects of language representation learning in terms of linguistic units of different granularity and in terms of unit selectivity and temporal dynamics, iv) comparison of alternative neural model architectures within the same experimental context, and v) comparing learning and representation extraction from both synthetic and real speech. In addition, we propose a new objective technique to evaluate the semantics learned by the audiovisual networks.

## Theoretical Background

One of the key challenges in early language acquisition research is to identify the fundamental computational principles responsible for the learning process. Young learners have to solve an apparently large number of difficult problems ranging from unit segmentation and identification to syntactic, semantic, and pragmatic learning on their way to become proficient language users. Is thereby unclear what type of collection of innate biases, constraints, and learning mechanisms are needed for language learning to succeed. In terms of parsimony, a theory should aim to explain the different aspects of LA with a minimal number of distinct learning mechanisms.

The key idea behind LLH is to replace several separate language learning processes and their proximal learning targets with a single general overarching principle for learning, namely predictive optimization. In short, LLH relies on the idea that the mammalian brain has evolved to become an efficient uncertainty reduction (=prediction) device, where input in one or more sensory modalities is used to construct a set of predictions regarding the overall state of the present and future sensorimotor environment (cf., e.g., Friston, 2010; Clark, 2013). This strategy has several ecological advantages. For instance, complete sensory sampling of the environment would take excessive time and effort, and actions often need to be taken with incomplete information of a constantly changing environment. In addition, predictive processing allows focusing of attentional resources on those aspects of the environment that have high information gain to the agent (see, e.g., Kakouros et al., 2018, for a review and discussion). As a result, the ability to act based on partial cues of the "external world state" (also across time) results in a substantial ecological advantage. Importantly, predictive processing necessitates some type of probabilistic processing of the stochastic sensory environment. This is because evaluation of the information value of different percepts requires a model of their relative likelihoods in different contexts (or degrees of "surprisal"; see also the Goldilocks effect; Kidd et al., 2012). This connects the overarching idea of predictive processing to the concept of *statistical learning* in developmental literature, as infants appear to be adept learners of temporal (Saffran et al., 1996) and cross-modal probabilistic regularities (Smith & Yu, 2008).

In the context of LLH, we postulate that statistical learning is a manifestation of general sensorimotor predictive processing, and where language learning could also be driven by optimization of predictions *within* and *across* sensory modalities[1] during speech perception. In order to efficiently translate heard acoustic patterns to their most likely visual referents or to predict future speech input, intermediate latent representations that best support this goal are needed. More specifically, the question is whether representation of the *linguistic structure* underlying the variable and noisy acoustic speech could emerge as a side product of such a predictive optimization problem (see also van den Oord et al., 2018).

In case of audiovisual associative learning, this idea can be illustrated by a simple high-level mathematical model such as

$$\arg_\theta \max p(\bar{v}_t \mid \bar{x}_t, \theta) \mid \forall t \in [0, T] \qquad (1)$$

where $\bar{v}_t$ is visual input at time *t*, $\bar{x}_t = \{x_0, x_1, x_2, \ldots, x_t\}$ is the speech input up to time *t*, $\theta$ is a statistical model (or biological neural system) enabling evaluation of the probability, and *T* is the total cumulative experience ("age") of the learner so far. Now, assuming that 1) $\theta$ consists of several *plastic* processing stages/modules $\theta = \{\theta_1, \theta_2, \ldots, \theta_N\}$ (e.g., layers or cortical areas in in artificial or biological neural networks), 2) Eq. (1) can be solved or approximated using some kind of learning process, and that 3) observed speech and visual input are statistically coupled, $\theta$ must result in intermediate representations that together lead to effective predictions of the corresponding visual world, given some input speech. If a solution for $\theta$ is discovered, i.e., *the model has learned to relate speech to visual percepts,* we can ask whether the intermediate stages of $\theta$ have become to carry emergent representations that correlate with how linguistics would characterize the structure of speech. Alternatively, if the model becomes able to understand even basic level semantics between speech and the visual word without reflecting any known characteristics of spoken language, that would be a curious finding in itself.

[1] In the most generic form also including motor aspects with articulation and manual gestures.

The basic formulation in Eq. (1) can be extended to model the full joint distribution $p(\bar{x}_t, \bar{v}_t|\theta)$ of audiovisual experiences. Alternatively, assuming stochasticity of the environment, it can be reformulated as minimization of Kullback-Leibler divergence between $p(\bar{v} \mid \psi)$ and $p(\bar{v} \mid \bar{x}, \theta)$, where $\psi$ is a some kind of stochastic generator of visual experiences (due to interaction with the world) and the latter term is the learner's model of visually grounded speech. However, the main implication of each of these models stays the same: discovering a model $\theta$ that provides an efficient solution to the cross-modal translation problem between spoken language and other representations of the external world. The same idea can be applied to within-speech predictions across time by replacing $\bar{v}$ with $\bar{x}_{t+k}$ ($k > 0$) in Eq. (1). In this case, if $k$ is set sufficiently high, the learned latent representations must generalize across phonemically irrelevant acoustic variation in order to generate accurate predictions for future evolution of the speech signal given speech up to time $t$; evolution which is primarily governed by phonotactics and word sequence probabilities in the given language (see van den Oord et al., 2018, for phonetic feature learning with this type of approach; cf. also models of distributional semantics, such as Mikolov et al., 2013, that operate in an analogous manner with written language).

Given the existence of modern deep neural networks, LLH can be investigated using flexible hierarchical models that can tackle complicated learning problems with real-world audiovisual data, and without pre-specifying the representations inside the networks. This is also what we do in the present study. While such computational modeling cannot tell us what exactly is happening in the infant brain, it allows us to investigate the fundamental feasibility of LLH under controlled conditions in terms of learnability proofs.

Note that we wish to avoid taking any stance on the debate whether discrete linguistic units are something that exist in the human minds or computational models. In contrast, we adopt a viewpoint similar to Ramscar and Port (2016) and use linguistic structure as an idealized description of speech data, investigating how the representations learned by computational models correlate with the manner that linguistics would characterize the same input. In addition, we do not claim that *audiovisual* learning is necessarily the only mechanism for early acquisition of primitive linguistic knowledge. We simply want to study the extent that this type process can enable or facilitate language learning, and generally acknowledge that purely auditory learning is also central to language learning.

## Earlier Related Work

A number of existing computational studies and machine learning algorithms have studied the use of concurrent speech and visual input to bootstrap language learning from sensory experience. In the early works (e.g., Roy and Pentland, 2002; Ballard and Yu, 2004; Räsänen et al., 2008; ten Bosch et al., 2008; Driesen and Van hamme, 2011; Yu et al., 2005; Mangin et al., 2015; Räsänen and Rasilo, 2015), visual information has been primarily used to support concurrent word segmentation, identification, and meaning acquisition. The basic idea in these models has been to combine cross-situational word learning (Smith & Yu, 2008)—the idea that infants learn word meanings by tracking co-occurrence probabilities of word forms and their visual referents across multiple learning situations—with simultaneous "statistical learning" of patterns from the acoustic speech signal. In parallel, a number of robot studies have investigated the grounding of speech patterns into concurrent percepts or actions (e.g., Salvi et al., 2012; Iwahashi, 2003). However, the acoustic input of some studies has been pre-processed to phoneme-like features (Roy and Pentland, 2002; Ballard and Yu, 2004; Salvi et al., 2012) or word segments (Salvi et al., 2012) using supervised learning. Alternatively, visual input to the models have been rather simplified, such as simulated categorical symbols for visual referents e.g., ten Bosch et al., 2008; Räsänen and Rasilo, 2015; Driesen and Van hamme, 2011).

In terms of LLH, the older models have had relatively rigid and flat representational structure, limiting their capability to produce emergent hierarchical representations. In contrast, the older models contain a series of signal processing and machine learning operations to solve the audiovisual task, including initial frame-level signal representation steps such as phoneme recognition or speech feature clustering, followed by pattern discovery from the resulting representations using transition probability analysis (Räsänen et al., 2008; Räsänen & Rasilo, 2015), non-negative matrix factorization (ten Bosch et al.,2008; Mangin et al., 2015), or probabilistic latent semantic analysis (Driesen & Van hamme, 2011), to name a few. Despite these limitations, these studies already demonstrate that access to units such as phonemes or syllables is not required for early word learning, as long as the concurrent visual information is related to the speech contents systematically enough. In addition, they show that word segmentation is not required before meaning acquisition, but that the two processes can take place simultaneously with referential meanings actually defining word identities in the speech stream. Such models can also account for a range of behavioral data from infant word learning experiments using auditory and audiovisual stimuli (Räsänen & Rasilo, 2015).

More recent developments in deep learning have enabled more advanced and flexible hierarchical models that can tackle richer visual and auditory inputs with unified elementary processing mechanisms. These models have their origins in methods for learning relationships between images and natural language descriptions of them, such as photographs and their written labels or captions (e.g., Frome et al., 2013; Socher et al., 2014; Karpathy & Li, 2015). These text-based models have been expanded to deal with acoustic speech input, such as spoken image captions (Synnaeve et al., 2014; Harwath and Glass, 2015; Harwath et al., 2016; Chrupała et al., 2017). Early works applied separate techniques for segmenting words-like units prior to alignment between audio caption data and images e.g. Synnaeve et al., 2014; Harwath and Glass, 2015). The more recent audiovisual algorithms operate without prior segmentation by mapping spoken utterances and full images to a shared high-dimensional vector space (Harwath et al., 2016; Chrupała et al., 2017). However, compared to text, dealing with acoustic speech data is a more difficult task: time-frequency structure of speech is not invariant similarly to orthography, but varies as a function of many different factors ranging from speaker identity to speaking style, ambient noise, or recording setup/listener situation. Moreover, acoustic forms of the elementary units such as phonemes or syllables are affected by the linguistic context in which they occur, causing substantial variation also within otherwise controlled speaking conditions. These are also challenges that language learning infants face, and which cannot be studied with transcription- or text-based models.

In a typical visually grounded speech (VGS) model (Harwath et al., 2016; Chrupała et al., 2017; see Fig. 1 for an example), the model consists of a deep neural network with two separate branches for processing image and speech data: an image encoder responsible for converting pixel-level input into high-level feature representations of the image contents, and a speech encoder doing the same for acoustic input. Both branches consist of several layers of convolutional or recurrent units, and outputs from the both branches are ultimately mapped to a shared high-dimensional semantic space, aka. *embedding space*, via a ranking function. The idea is to learn neural representations for images and spoken utterances so that the embeddings produced by both branches are similar when the input images and speech share semantic content. Once trained, distances between the embeddings derived from inputs can then be used for audiovisual, audio-to-audio, or visual-to-visual search, such as finding the semantically best matching images for a spoken utterance, or finding utterances with similar semantic content than a query utterance (Harwath et al., 2016; Chrupała et al., 2017; see also Azuh et al., 2019, and Ohishi et al., 2020, for cross-lingual approaches).

Training of these models is carried out by presenting the network with images paired with their spoken descriptions (whose mutual embedding distances the model tries to minimize) and pairs of unrelated images and image descriptions (whose embedding distances the model tries to increase). The visual encoder is often pre-trained on some other dataset using supervised learning (but see also Harwath et al., 2018), whereas the speech encoder and mappings from both encoders to the embedding space are optimized simultaneously during the training. Model training is typically conducted on datasets specifically designed for the image-to-speech alignment tasks, either by adding synthesized speech to captioned image databases, such as SPEECH-COCO by Havard et al. (2017) or Synthetic Speech COCO (SS-COCO; Chrupała et al., 2017) derived from images and text captions of MS-COCO (Chen et al., 2015), or acquiring spoken descriptions for images using crowd-sourcing, such as Places Audio Caption Corpus (Harwath et al., 2016) derived from Places image database (Zhou et al., 2014) or SpokenCOCO (Hsu et al., 2020) derived from MSCOCO.

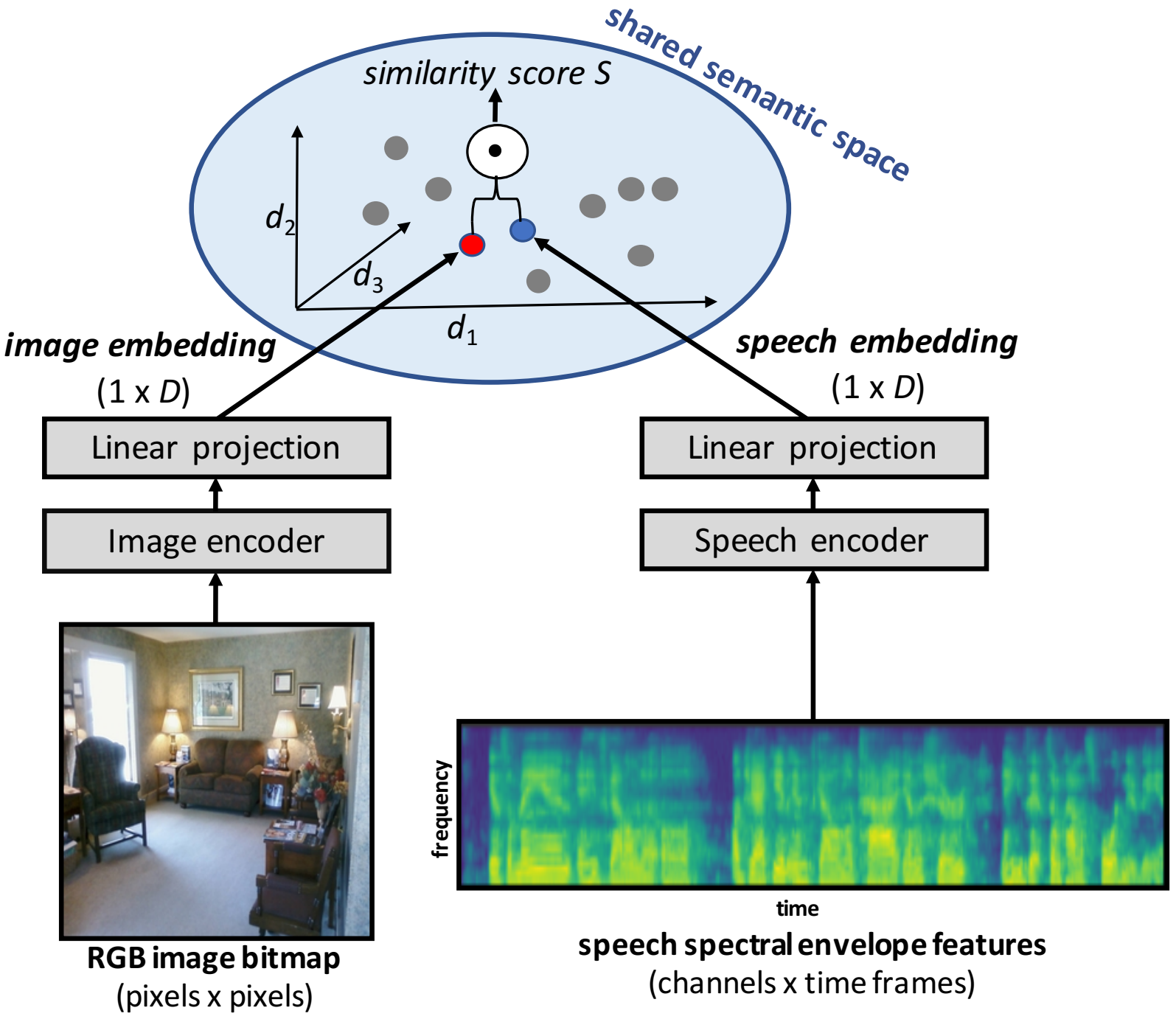

**Figure 1.** ***The basic architecture of the VGS models explored in the present study. Visual and auditory input data are processed in two parallel branches, both consisting of several neural network layers. Outputs from both branches are mapped into a shared "amodal" embedding space that encodes similarities shared by the two input modalities.***

### *Evidence for Language Representations in VGS Models*

From the perspective of LLH, the question of interest is whether the audiovisual models learn latent representations akin to linguistic structure of speech, as the models learn to map auditory speech to semantically relevant visual input and vice versa. In this context, a number of studies have investigated phonemic learning in VGS models.

Alishahi et al. (2017) used a recurrent highway network (RHN)—a variant of recurrent neural network (RNN)—VGS model with 5 recurrent layers to investigate how phonological information is represented in intermediate layers of the model (same model as used by Chrupała et al., 2017). Using synthetic speech from SS-COCO, they trained supervised phone classifiers with input-level Mel-frequency cepstral coefficients (MFCCs) and hidden layer activations as features to test how informative the features are with respect to phonetic categories. Alishahi et al. found that, even though the MFCCs already led to approximately 50% phone classification accuracy, the accuracies improved substantially when using activations from the first two recurrent layers of their model (up to approx. 77.5%) and then decreased slightly for the last recurrent layers. To further probe phonetic and phonemic nature of their network representations, Alishahi et al. (2017) also applied a so-called minimal-pair ABX-task (Schatz et al., 2013) to the networks to test whether the hidden representations can distinguish English minimal pairs in speech. Again, the best phonemic discriminability was obtained for the representations of the first two recurrent layers. Alishahi et al. (2017) also applied agglomerative clustering to activations within each layer, and found that the pattern of feature organization in MFCCs and in the first recurrent layer were better correlated with the ground-truth phoneme categories than the activations computed from other layers.

Drexler and Glass (2017) also used the ABX-task to investigate phonemic discriminability of the hidden layer activations of a CNN-based VGS model from Harwath and Glass (2017). Similar to Alishahi et al. (2017), they found that the hidden layer activations were better than the original spectral input features in the ABX-task (among other tasks), that the early layers were phonemically more informative than the deeper ones, and that the network also learned to discard speaker-dependent information from the signal due to the visual grounding. However, they also found that somewhat higher phonemic discriminability was still obtained using purely audio-based unsupervised learning algorithms compared to their VGS model. Another study by Harwath et al. (2020) augmented the CNN-based VGS model from Harwath et al. (2018) with automatic discretization (vector quantization) of the internal representations during the training and inference process. Then they investigated how this affects the phonemic and lexical discriminability of the hidden layer representations. They found that phoneme discrimination ABX scores of the early layer representations were much higher than those typically observed for spectral features in the same task or with a number of baseline speech representation learning algorithms. They also found that discretized representations from early layers primarily carried phonemic information, while representations quantized in deeper layers corresponded better to lexical units. However, discretization did not improve phonemic discriminability beyond the original distributed multivariate representations of the hidden layers.

Recently, Räsänen and Khorrami (2019) trained a weakly supervised convolutional neural network (CNN) VGS model to map acoustic speech to the labels of concurrently visible objects attended by the baby hearing the speech, as extracted from head-mounted video data from real infant-caregiver interactions of English-learning infants (Bergelson & Aslin, 2007). They then measured the so-called phoneme selectivity index (PSI) (Mesgarani et al., 2014) of the network nodes and layers. Their results indicated that, in addition to learning a number of words and their referents from such data, hidden layer activations of the model also became increasingly representative of phonetic categories towards deeper layers of the network. The model was also able to handle referential ambiguity in the visual input when the infant was not attending the correct object. However, Räsänen and Khorrami did not use actual visual inputs but categorical labels of the perceived objects, simplifying the visual recognition process substantially.

In terms of phone segmentation, Harwath and Glass (2019) investigated whether activation dynamics of a CNN-based VGS model reflect underlying phonetic structure of speech. They compared temporal activation patterns of VGS-model hidden layers to phone boundaries underlying the input speech data from TIMIT corpus (Garofolo et al., 1993). As a result, they found that peaks in the change-rate of activation magnitudes of the early CNN layers were highly correlated with transitions between phone segments. In contrast to studying whether the models learn to segment, Havard et al. (2020) studied how the performance of VGS models improves if linguistic unit segmentation is provided as side information to the model during the training. They found that explicit introduction of segmentation cues led to substantial performance gains in the audiovisual retrieval task compared to regular VGS training. The effect was the most pronounced when the system was supplemented with a hierarchy of phone, syllable, and word boundaries across different layers of the model.

Several studies have also investigated lexical representations in VGS-based models. Chrupała et al. (2017) used the same RHN-RNN networks as Alishahi et al. (2017) and showed that the RHN model outperformed the earlier CNN model of Harwath et al. (2016) on audio-to-image retrieval task. Then they investigated how linguistic form- and semantics-related aspects of the input are encoded in the hidden layers of the network. Through a number of experiments, Chrupała et al. (2017) showed that form related features become represented within the first layers of their model, whereas deeper layers tended to encode semantics better than the early layers. They also studied how the network responds to homonyms (i.e., words with similar pronunciation but different meaning, such as "*sail*" and "*sale*") and concluded that the representations of deeper network layers became increasingly better at distinguishing homonyms. In other words, the deep representations also contained cues for contextual semantic disambiguation.

Harwath and Glass (2017) investigated whether word segments in speech can be connected to the bounding boxes of corresponding objects in images using a convolutional neural model of VGS, and showed that this was indeed the case. As an extension to their work, Harwath et al. (2018) created a method to map segments of spoken utterances to their associated objects in the pictures (referred to as "match-map" network) in order to investigate how object and word localization emerges as a side-product of training a network using caption-image pairs. In another study, Havard et al. (2019b) studied if lexical units can be segmented from the representations of recurrent layers of a RNN-based VGS model. By using a variety of metrics, they showed that the network learns an implicit segmentation of word-like units and manages to map individual words to their visual referents in the input images.

Kamper et al. (2017) have also studied if visual data can be employed as an auxiliary intermediate tool for detecting words within speech signals. They designed a speech tagging algorithm which is trained using a dataset of aligned speech-image pairs. They first trained a supervised vision tagging system which, given an image, generates probabilities for the presence of different objects within that picture. Next, they integrated their trained vision model with an audio processing network and trained a joint system which maps spoken utterances to the visual object probabilities. As a result, their network learned to output a list of keywords (object category names) given continuous speech input, again without ever receiving direct information on what constitutes a word in an acoustic sense.

Merkx et al. (2019) further improved the audiovisual search performance of the RNN-based VGS model of Chrupała et al. (2017) and used it to study how different layer activations of the model encode words in speech. They used acoustic input features and hidden layer activations as inputs to a supervised word classifier to test if the representations are informative with respect to underlying word identities. They concluded that the presence of individual words in the input can be best predicted using activations of an intermediate (recurrent) layer of their model.

Havard et al. (2019a) studied neural attention mechanism (Bahdanau et al., 2015) in an RNN-based VGS model using English and Japanese speech data. They found that similar to human attention (Gentner, 1982), neural attention mostly focuses on nouns and word endings. This is in line with the knowledge that infant early vocabulary tends to predominantly consist of concrete nouns. In another study, Havard et al. (2019b) examined the influence of different input data characteristics in a word recognition task by feeding the VGS model with synthesized isolated words with varying characteristics. They observed a moderate correlation between word recognition accuracy and frequency of the words in training data, and a weak correlation for image-related factors such as visual object size and saliency. Havard et al. (2019b) also investigated word activations in the same RNN model using the so-called gating paradigm from speech perception studies (Grosjean, 1980). For this purpose, they fed the network with individual spoken words and truncated the words from different positions at the beginning or end of the words. They found that the precision of word recognition dropped steeply if the first phoneme of a word was removed. In contrast, removal of the word-final phonemes had little impact on precision, and the precision decreased steadily when truncating additional phonemes from the end. This was generally in line with data from human lexical decision tasks.

Inspired by the work of Havard et al. (2019b), Scholten et al. (2020) recently studied word recognition in an RNN-VGS model. Instead of using synthesized speech, they conducted their experiments using real speech data from Flickr8k (Harwath & Glass, 2015). Scholten et al. evaluated their model on word recognition by examining how well word embedding vectors can retrieve images with the correct visual object corresponding to the query word, measuring the impact of different factors on word recognition performance. They found that longer word lengths and faster speaking rates were negatively correlated with performance, while word frequency in the training set had a substantial positive impact on the task performance.

Overall, the general finding from the earlier work has been that the representations learned by VGS models exhibit many characteristics related to the underlying linguistic structure of the input speech, and they learn this structure without ever receiving specifications of how speech or language are organized into some kind of elementary units. This suggests that phonetic and lexical representations and segmentation capabilities could emerge as a side-product from meaning-driven learning. However, it is not yet clear in which conditions these phenomena can occur, and how different levels of language representation are related to each other inside the same models. This is since the studied model architectures (RNNs vs. CNNs), model analysis methods (discriminability, clusteredness, node vs. layer selectivity etc.), and data (synthetic vs. real speech) utilized by the previous studies have varied from one study to another. No individual study has attempted to look at the emergence of linguistic units at phonetic, syllabic, and lexical levels in a single model or study, nor compared multiple model architectures within the same experimental context. In addition, the existing studies have rarely reported baseline measures from untrained models, making it unclear how much of the findings are actually driven by the visually-guided parameter optimization compared to the effects of non-linear network dynamics also present with randomly initialized model parameters (see also Chrupała et al., 2020). This leaves unclear questions such as: 1) Can a single neural model reflect emergence of several levels of linguistic structure at the same time, including phone(me)s, syllables, and words, both in time and selectivity? 2) If so, does the network encode such units preferentially in terms of individual selective nodes or distributed representations? 3) How robust these findings are to variations in the neural architecture of VGS models? 4) Do the analysis findings (primarily carried out on synthetic speech) also generalize to real speech with higher acoustical variability?

In our experiments, we seek to address the above questions by systematically investigating the audiovisual aspect of LLH in three alternative VGS network architectures and at phoneme, syllable, and word level, both in terms of selectivity and in terms of temporal characteristics, and using both synthetic and real speech datasets. The second section describes the alternative speech processing networks used in our experiments, followed by methodology to analyze the internal representations of the models with respect to linguistic structure underlying the speech input to the model. In the third section, we describe the data and experimental setup of our study, followed by results, discussion, and conclusions.

## Methods

The goal of our experiments was to investigate the extent that linguistic units of different granularity may emerge as a side product of audiovisual cross-situational learning in neural models of visually grounded speech. We also study the extent that the architecture of the model or type of data (real vs. synthetic) affects the nature of the learned representations.

We first explain the adopted VGS model structure in more detail, including three alternative speech encoder architectures explored in our experiments. We then describe our toolkit used to analyze the hidden layer representations of the networks with respect to linguistic characteristics of the input speech. In addition, we propose a new automatic method for evaluating the semantic relevance of the audiovisual associations learned by the models.

### Model Architecture and Speech Encoder Variants

VGS systems are generally trained to align between speech and image modalities so that they learn semantic similarities between the two modalities without any explicit supervision in the form of labels. Here our aim is to use VGS models to simulate infants' audiovisual learning, where they hear speech that is related to the observable visual contexts, but does not contain unambiguous and isolated speech-referent pairs. The setup thereby simulated cross-situational word learning under a high degree of referential uncertainty, and without access to prior segmentation of acoustic word forms.

We follow the methodology by Harwath and Glass (2017) and Chrupała et al. (2017), where input to the model consists of images (photographs) and their spoken descriptions. Speech and image data are initially processed in different encoders consisting of several ANN layers, followed by encoder-specific mappings to a shared "amodal" embedding space. In this space, a chosen similarity metric can be used to measure the pairwise similarity of any representations resulting from auditory or visual channels. During training, the model is optimized to assign a higher similarity score for embeddings resulting from images and image descriptions that match with each other (so-called *positive samples*). At the same time, the model tries to assign higher distances for embedding pairs from unmatched images and utterances (*negative samples*). As a result, the model learns to generate embeddings that encode concepts available in both input modalities. The basic architecture of the image-to-speech mapping network is shown in Fig. 1.

In our current visual encoder network, pixel-level RGB image data are first resampled to 224x224 pixels and then transformed into high-level features using VGG16 image classification network (Simonyan and Zisserman, 2015), which is a deep CNN consisting of 16 layers pretrained on ImageNet data (Russakovsky et al., 2015). Output features of the first fully connected layer (14th layer) of VGG16 are then projected linearly to a $D$-dimensional space to form the final visual embeddings, and where the linear layer weights are optimized during the VGS model training.

### *Compared Speech Encoder Architectures*

We compare three alternative speech encoder networks, all consisting of a stack of convolutional and/or recurrent neural layers applied on speech input. In all models, the input speech is represented by 40-dimensional log-Mel filterbank energies extracted with 25-ms windows with 10-ms window hop-size, which is a representation that simulates the frequency-selectivity of the human ear. The following three speech encoder architectures were investigated in our experiments (Fig. 2):

**CNN0** (Fig. 2, left) is a multi-layer convolutional network with an architecture adopted from Harwath and Glass (2017). It includes five convolutional layers with increasing temporal receptive fields, each followed by a max pooling layer. The output of the last convolutional layer is pooled over the entire utterance in order to discard the effects of absolute temporal positioning of the detected patterns.

As an alternative convolutional model, we designed a **CNN1** network (Fig. 2, middle) with 6 convolutional layers and hand-crafted receptive field time-scales in different layers. We specified the convolutional and pooling layers such that the filter receptive field sizes at different layers would approximately correspond to the known typical time-scales of phones, syllables, and words while gradually expanding towards the larger units (see Fig. 2 for details). As in CNN0, the output of the last convolutional layer is maxpooled across all the time steps.

Our third model variant, **RNN** (Fig. 2, right), was adapted from the model introduced originally by Chrupała et al. (2017) and also used by Alishahi et al. (2017). It includes a convolution layer as the first layer, followed by three residualized recurrent layers with Long Short-Term Memory (LSTM) units. Unlike Chrupała et al. (2017), we use three layers instead of the original five layers, as we observed in our initial tests that the three layer model was already capable of achieving comparable performance to the CNN models in the audiovisual mapping task while training much faster than the original model. Also, in order to maintain comparability of the three networks, we do not utilize a separate attention mechanism in the RNN model. The first two recurrent layers of the RNN feed their frame-by-frame activations to the next layer, allowing measurement of their temporal activations. In contrast, the last layer outputs an activation vector for the entire test sentence after processing it fully, discarding the frame-based temporal information.

In all three variants, the utterance-level activations of the final layer are L2 normalized and linearly projected to $D$-dimensional latent space to form the final speech embeddings. These can then be compared to other embeddings within and across the modalities. We use cosine similarity to measure a similarity score $S$ between any two embeddings.

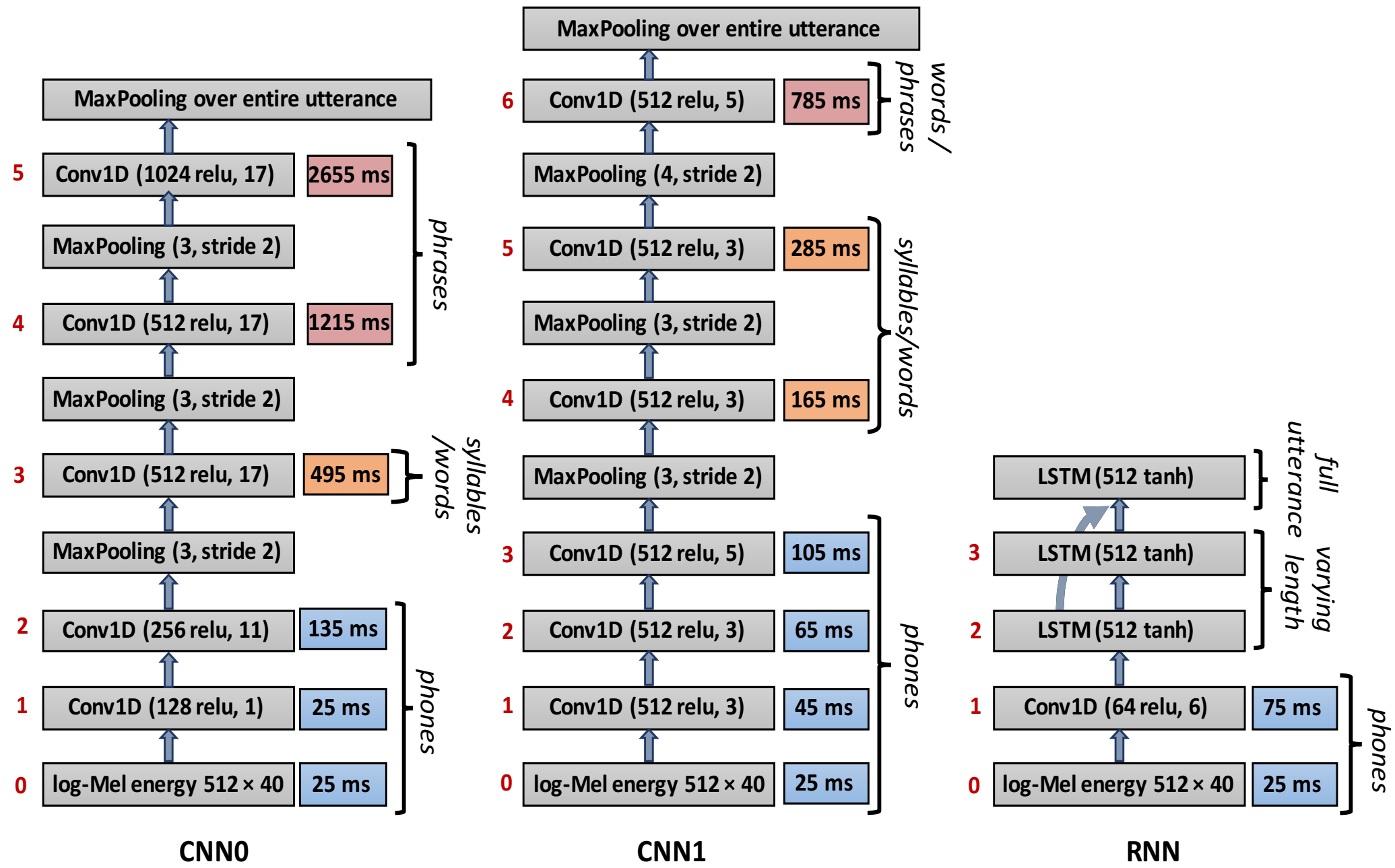

**Figure 2.** ***Three speech encoders studied in our experiment together with the maximum temporal receptive field lengths of the network nodes. Left: CNN0. Middle: CNN1. Right: RNN. Unit descriptions next to the layers denote the approximate linguistic unit time-scale that the receptive fields of the convolutional layers correspond to. Numbers in red denote layer identifiers used in the analyses of section Results.***

Note that both the CNN and RNN -based models are capable of modeling temporal structure of the data. On one hand, recurrent layers are specifically designed for processing sequential data because they can potentially memorize the history of all past events and therefore recognize patterns across time. On the other hand, convolutional layers are also capable of capturing temporal structure through the hierarchy of increasingly large temporal receptive fields (Gehring et al., 2017), where the largest receptive field size also sets the limit on the temporal distance up to which they can capture statistical dependencies in the data. However, the manner that CNNs and RNNs models capture the temporal structure is very different. Therefore it was of interest whether we can see commonalities or differences in their strategy of encoding linguistic structure of the speech data in order to solve the audiovisual mapping problem.

### *Model Training*

The method we applied for training our networks followed the same strategy as in Harwath et al. (2016) and Chrupała et al. (2017) by using the so-called *triplet loss*: first, a triplet set is made by taking one matching image-speech pair (i.e., an image and an utterance describing it), and adding two negative samples by pairing the original image with a random speech utterance and the original utterance with a random image. The data are then organized into a collection of *B* such triplets. At training time, error backpropagation is used to minimize the following loss function:

$$L(\theta) = \sum_{j=1}^{B} \max\left(0, S_j^c - S_j^p + M\right) + \max\left(0, S_j^i - S_j^p + M\right) \qquad (2)$$

where $S_j^p$ is the similarity score of *j*th ground-truth pair $S_j^c$ the score between original image and the impostor caption, and $S_j^i$ is the score between original caption and the impostor image. In practice, the loss function decreases when ground-truth pair embeddings become more similar to each other. Similarly, the loss decreases when mismatched pairs get further away from each other until they reach distance of *M*, which is referred to as the *margin* of the loss. Intuitively, this means that when the embeddings of a false pair are more than *M* units apart, they are considered as semantically unrelated and the pair no longer affects further parameter updates of the model. As a result, the model learns to tell apart semantically matching and mismatching audiovisual inputs.

### Model Evaluation

Our model evaluation consisted of two stages. We first verified that the trained networks have successfully learned to associate auditory and visual patterns to each other, as measured in terms of semantic retrieval tasks. We then proceeded to analyzing whether and how the hidden layer representations of the models correlate with linguistic characteristics of the input speech. Methods and metrics for these analyses are described next.

### *Audiovisual Search Performance*

After training, audio and visual embedding layers can represent semantic similarities between images and spoken captions using the similarity score. Therefore, within a pool of test images and utterances, semantically related examples can be distinguished by sorting instances based on the mutual similarities between their embedding vectors. As a quantitative evaluation of model performance, we studied *recall@k* introduced by Hodosh et al. (2013) and frequently applied in VGS model literature. In the present case, recall@k measures performance of the trained models on image search, given an input utterance as a query ("speech-to-image search"), and on automatic image caption search, given an image as a query ("image-to-speech search", sometimes also referred to as automatic image annotation; see also Harwath et al., 2016 and Chrupała et al., 2017).

For measuring recall@k, spoken captions and images from a test dataset are presented to speech encoder and image encoder branches of the model, respectively, resulting in speech and image embedding vectors. In speech-to-image search task, the similarity of each speech sample with all test images is then calculated by applying a similarity metric (here: cosine similarity) to their embedding vectors, and $k$ nearest matches are maintained. Recall@k is then obtained as the percentage of utterances for which the image corresponding to the utterance is within the $k$ closest matches. Similarly, for image-to-speech search task, recall@k measures the percentage of query images for which the correct caption is within the $k$ closest retrieved utterances.

In our experiments, we report recall@10 as it is also commonly used in earlier studies (Harwath et al., 2016; Chrupała et al., 2017).

### *Quantitative Evaluation of Audiovisual Search Semantics*

While previous studies have primarily used recall@k to measure performance in audio-visual alignment tasks, the problem of recall@k is that it is unable to account for semantically relevant matches beyond the pre-defined image-caption pairs of the database (see Kamper et al., 2019). For instance, the data might contain a large number of food pictures, and hence a spoken query such as "There's leftover food on the table" could result in many relevant search results with food in them, but only the one for which the caption was originally created for would be counted as a correct search result. For this reason, Kamper et al. (2019) used human judgments for evaluating semantic retrieval in his VGS model. However, despite crowdsourcing, this can be time consuming and expensive.

In order to objectively evaluate and compare the quality of the learned semantic representations of the alternative speech encoder architectures, we developed a new method to objectively and automatically evaluate semantic similarity between input speech and the corresponding retrieved audio captions. For this purpose, we utilized Word2Vec (Mikolov et al., 2013) and SBERT (Reimers et al., 2019), distributional word semantics models trained on large-scale text data, that allow measurement of semantic similarity between different words (Word2Vec) or sentences (SBERT) in textual form. Since semantic similarity judgements of distributional semantic models correlate highly with human ratings of similarity and synonymity (Landauer and Dumais, 1997, or Günther et al., 2019, and references therein; but see also Nematzadeh et al., 2017 or Deyne et al., 2021, for recent analysis), we use these two models as proxies for human judgement for semantic relatedness between different spoken captions.

With SBERT[2], the semantic similarity of two captions can be obtained simply by taking the cosine similarity of the sentence-level embeddings extracted from the utterance transcripts. However, the maximum similarity score is strongly affected by presence of repeated words in the two compared sentences. An alternative measurement can be obtained by excluding repeating words between the sentences, but we hypothesized that removing of content words might cause unwanted problems with context-dependent embeddings of SBERT. In order to measure semantic similarity of two spoken captions at the word level, we first extracted content words of the utterance transcripts using the Natural Language Toolkit (NLTK) in Python by including nouns, verbs, and adjectives while ignoring other parts of speech. We then calculated semantic relatedness score (SRS ∈ [0, 1]) between the two utterances as:

$$SRS(reference, candidate) = \frac{1}{N_r} \sum_{i=1}^{N_r} \max\{S_{\mathrm{w2v}}(r_i, c_j) \mid \forall j\} \quad (3)$$

where $S_{\mathrm{w2v}}$ is Word2Vec similarity score between individual words (cosine similarity of the pre-trained word embedding vectors) and $r$ and $c$ are content words in reference and candidate sentences, respectively. In other words, for each content word in the reference utterance, the most semantically similar word is chosen from the candidate utterance, and the total similarity score is the average across all such pairings. By excluding the repeating words between the sentences before SRS calculation, this measurement is then an indicator of semantic relatedness of the utterances while ensuring that the similarity is not simply driven by identical lexical content. In our experiments, we used both SBERT and SRS semantic similarity measurements to test

[2] We used pre-trained SBERT model "paraphrase-distilroberta-base-v1" trained on paraphrase data (Reimers and Gurevych, 2020)

whether audio-to-audio search results produce semantically meaningful outputs even if the utterances do not correspond to the same original image, thereby enabling more representative evaluation of semantic retrieval beyond recall@k.

***Selectivity Analysis of Hidden Layer Activations***

The literature on interpreting linguistic structure learned by deep neural networks has shown that multiple alternative metrics are needed to understand hidden representations. This is since there is no unanimous view of what "linguistic representations" should look like in such a distributed multi-layer representational systems, and hence it is difficult to operationalize broad concepts such as as "phonemic or lexical knowledge" in terms of specific and sensitive measures to probe the hidden layer activations (see, e.g., Belinkov & Glass, 2020; Chrupała et al., 2020). Given this starting point, our metrics for analyzing the relationship between model activation patterns and linguistic units in the speech input focus on four complementary measures: selectivity of individual nodes in network layers towards specific linguistic units, clusteredness of entire activation patters of a layer, and linear and non-linear separability of layer activations w.r.t. different linguistic unit types. We deliberately focus on statistical and classifier-based measures of analysis that are suitable for basic level categorical data (phone, syllable, or word types), whereas measures such as representational similarity analysis (RSA; Kriegeskorte et al., 2008) used in some other works (e.g., Chrupała et al., 2020) are better suited for non-categorical reference data[3].

This section uses phones as the example units of analysis, but the same analysis process was also carried out for syllables and words in each layer of each of the compared models, as described in section Model Evaluation. As is customary, we use *types* to refer to unique phones in the corpus and *tokens* for individual occurrences of phones in the data.

The first measure, ***node separability***, describes how well activations corresponding to the different phones in the speech input can be separated by individual nodes of a layer. The metric is based on d-prime measure (aka. sensitivity index) from the signal detection theory. While standard d-prime describes the separation of two normal distributions in terms of how many standard deviations (SDs) their means are apart, *D*-dimensional generalization of the metric can be written as:

$$\bar{d}'_{i,j} = \frac{\bar{\mu}_i - \bar{\mu}_j}{\sqrt{\frac{1}{2}(\sigma_i^2 + \sigma_j^2)}} \qquad (4)$$

where $\bar{\mu}_i$ and $\bar{\mu}_j$ indicate the means and $\sigma_i^2$ and $\sigma_j^2$ SDs of the *D*-dimensional activations (of a layer with *D* nodes) during specific phones $i, j \in \{1, 2, \ldots, M\}$, respectively. By taking the root-mean-square of across the *D* nodes and then averaging the result across all possible unique pairs of phones, we obtain the multidimensional node separability measure $d' \in [0, \infty]$ for the given layer:

$$d' = \frac{2}{M^2 - M} \sum_{i=1}^{M-1} \sum_{j=i+1}^{M} \sqrt{\frac{1}{D} \sum (\bar{d}'_{i,j})^2} \qquad (5)$$

[3] The main advantage of RSA is its sensitivity to different grades of similarity between the analyzed entities. However, derivation of reference metrics for linguistic representations could be conducted in various ways, including factors such as phonotactics or articulatory attributes for phones, focusing on semantics, syntactic role, or lexical neighborhood density for words, or using human similarity judgements or brain imaging data for any of the units. Different choices on the relative importance of such factors could also lead to different analysis findings.

The metric is independent of representation space dimensionality. It is zero if all nodes have identical activation distributions for all phone types, and grows with increasing separation of the distributions for different phone types. Intuitively, if individual nodes of a layer specialize in encoding different phone categories, we should observe a high value of *d'* for the given layer.

Our second measure investigates the degree that the distributed activation pattern of an entire layer encodes phonetic identity. We measure this ***clusteredness*** of the representations by applying *k*-means clustering to the extracted activations of each layer, where the number of clusters *k* is specified to be the same as the number of phone types in the corpus (i.e., $k = M$; see also Alishahi et al., 2017, for an agglomerative approach). Clustering is initialized randomly, and then all activation vectors get assigned to one of the clusters by the k-means algorithm. The proportions of samples from each phone type in each cluster are then calculated, and each cluster is assigned to represent a unique phone type. The assignment is based on greedy optimization, where the cluster with the highest proportion of samples from a single phone category (i.e., having the highest phone *purity*) is chosen as a representative of that type, and then that cluster and phone type are excluded from the further assignments. The process is repeated until all clusters have been mapped to their best-matching types (with the aforementioned constraints). The overall phonetic purity of the clustering is then measured as the average of the cluster-specific purities w.r.t. to the assigned phone categories. The result is averaged across 5 independent runs of k-means to account for the variance due to the random initialization. Mean and SD of the overall purity across the runs are then reported in the experiments. Purity ranges from 1/*M* (different phones are uniformly distributed across all clusters) to 1 (phones group into perfectly pure clusters in an unsupervised manner).

Besides analyzing the activations of individual nodes and full layers, we use two additional measures to investigate whether the full layers or their node subsets separate between different phone types: ***linear separability*** and ***non-linear separability***, as measured by machine learning classifiers that are trained to classify phones using the activation patterns as features (also known as *diagnostic classifiers*; see also Belinkov & Glass, 2020; Chrupała et al., 2020). For linear separability, we use support vector machines (SVMs) with a linear kernel. For non-linear separability, we use a k-nearest neighbors (KNN) classifier. Both classifiers are trained with a large number of phone tokens from each phone type, and then tested on held-out tokens from the same types (see section Model Evaluation for details). Separability is measured in terms of unweighted average recall (UAR %), corresponding to the average of phone-specific classification accuracies.

On top of the four reported metrics, we also calculated a number of other metrics. For the node selectivity, we measured the so-called Phoneme Selectivity Index (PSI) by Mesgarani et al. (2014). Since PSI was very highly correlated with the d-prime separability across the different layers and test conditions, we do not report it separately. In addition, we measured the difference and ratio between cross- vs. within-type cosine distances of layer activation vectors as a measure of separability. However, we found the k-means-based metric more representative and straightforward to interpret for the phenomenon of interest. Finally, we also calculated overall classification accuracies (aka. weighted average recall / WAR) for the SVM and KNN classifiers. Since WAR is simply the proportion of tokens correctly classified, it is biased towards classification accuracy of more frequent phones. However, UAR and WAR were also highly correlated, and therefore we report UAR only.

In addition, we initially performed word-level analyses separately for content words only, as the we hypothesized that the audiovisual learning paradigm may support learning of nouns and verbs better than, e.g., function words. However, the results were highly correlated to those using all word types in the analyses. For the sake of clarity, we only report the results for words from all parts of speech.

***Temporal Analysis of Hidden Layer Activations***

We also compared temporal dynamics of the network activations with ground truth phone, syllable, and word boundaries. Our question was whether the temporal activation patterns would somehow reflect the underlying linguistic unit boundaries, i.e., whether the models reflect emergent speech segmentation capabilities even though they were not trained for such a purpose. In earlier work, Harwath and Glass (2019) reported that activation magnitudes of a VGS model (similar to our present CNN0) were related to phone boundaries on TIMIT corpus (Garofolo et al., 1993) after the model had been trained on Places Audio Caption Dataset (Harwath et al., 2016). Our present aim was to replicate the finding on other corpora, and to investigate segmentation of syllables and words in addition to phones.

In order to do so, we first measured activations of each layer for each input utterance as a function of time, and then characterized the overall temporal dynamics using a 1-D time-series representation for the given input. We then compared the peaks of this representation with known linguistic unit boundaries. We investigated three types of 1-D representations for the network temporal dynamics: activation magnitudes $m_l[t] \in [0, \infty]$ (from Harwath & Glass, 2019), instantaneous normalized entropy $h_l[t] \in [0, 1]$, and linear regression from instantaneous node activations to pseudo-likelihoods of unit boundaries, $r_l[t] \in [-\infty, \infty]$. The first one is simply the L2-norm of activations of all nodes $n$ in layer $l$ at time $t$. Entropy was defined as

$$h_l[t] = -\frac{\sum_{n=1}^{D} \hat{a}_n[t] \log_2(\hat{a}_n[t])}{\log_2(N)} \qquad (6)$$

where $\hat{a}_n[t]$ denotes the node- and layer-specific activations after the sum of activations has been normalized to 1 for each $t$ and $l$, and where $D$ is the total number of nodes in the given layer. In essence, $m_l[t]$ quantifies how well the input matches to the receptive fields of the filters in each layer, whereas $h_l[t]$ quantifies how the activity of the layer is distributed: small values close to zero indicate that only few neurons are active at the given time, whereas $h_l[t]$ close to 1 (high entropy) means that all nodes have very similar activation levels and hence little information is transmitted by the instantaneous activations.

Linear regression was performed by first creating a target temporal signal for each utterance, where the signal had a Gaussian kernel with a maximum amplitude of one centered at each unit boundary (see Landsiedel et al., 2011, for a similar approach for syllable nuclei detection). Duration of the kernels was set so that approximately 95% of the kernel mass was within ±20 ms from the annotated target boundary for each phone and within ±40 ms for syllables and words. This was done to account for the uncertainty in defining the exact unit boundary positions in time (see, e.g., Kvale, 1993). Then an ordinary least-squares linear mapping was estimated from the instantaneous node activations to the target signal. After estimating the mapping, the regression representation $r_l[t]$ was obtained by applying the mapping to all activations in the corpus, representing the estimated "score" that a boundary is located at each temporal position. A separate mapping model was trained for phones, syllables, and words to be used in respective evaluations. Due to computational constraints, a subsample of 2,500 target corpus utterances was always used to train the regression model.

The first two representations, $m_l[t]$ and $h_l[t]$, were normalized to have zero mean and unit variance at the utterance-level before further analysis. Due to the nature of the regression targets, $r_l[t]$ was already targeted between 0 and 1 (except for regression inaccuracies) and did not require further normalization.

Following Harwath and Glass (2019), a difference of a Gaussian filter of $\sigma = 5$ ms was applied to the normalized 1-D curves from L2-norm and entropy to measure their rate of change, followed by filter delay correction (see Fig. 3 for visualization). After preprocessing each of the 1-D representations, peak-picking was applied to detect local maxima in the rate of change in magnitude or entropy or maximum boundary score in the regression output. The outputs of the peak-picking were then considered as boundary hypotheses and compared to annotated linguistic unit boundaries. Sensitivity of the peak picking algorithm was controlled by a detection threshold $\theta_d$—the minimum required difference between the last local minimum and current local maximum in order for the maximum to be considered as a peak.

Phone segmentation was evaluated using standard metrics, where a reference phone boundary was considered as correctly detected if the algorithm had produced a hypothesized boundary within ±20 ms from the reference (Räsänen et al., 2009). Similar procedure was used for syllable and word segmentation but using a ±50-ms criterion for the detection, as used in the earlier literature on syllable segmentation (Räsänen et al., 2018, and references therein). Recall (proportion of boundaries detected), precision (proportion of hypothesized boundaries correct), and F-score (harmonic mean of the previous two) were then calculated as the primary metrics for segmentation.

For conciseness, we only report results for the optimal $\theta_d$ determined separately for phones, syllables, and words across the full test set. This is since we are primarily interested in whether the model activation patterns reflect boundaries of linguistic units, not whether the settings of our algorithm generalize to novel test conditions as required for proper speech segmentation algorithms. For the same reason, the linear regression model was trained on the same data as to which it was then applied to.

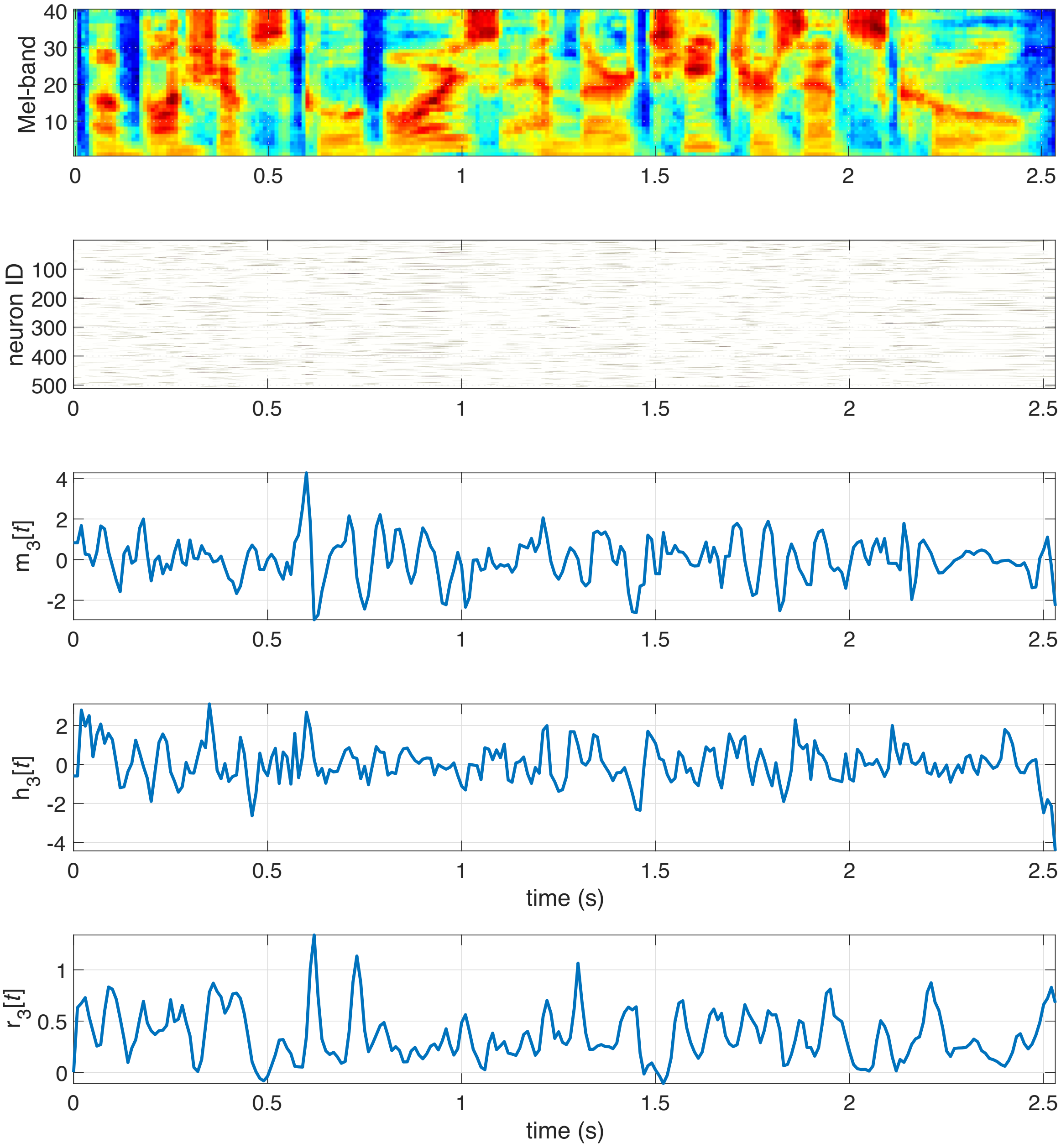

**Figure 3.** ***An example of the temporal analysis process for a spoken utterance. Top: original speech input as a log-Mel spectrogram. Second panel: corresponding neuron activations from layer 3 of the CNN1 model. Third panel: instantaneous magnitude of the node activations. Fourth panel: instantaneous entropy of the activations. Bottom panel: linear regression from the node activations to phone boundary scores. Segment boundaries are obtained from the curves with peak-picking. The first two activation curves are z-score normalized.***

## Experimental Setup

### *Data*

We investigated model training and representation analysis with both synthetic speech and real speech. The use of synthetic speech allows highly controlled experiments with clean signals, limited number of speakers, and accurate ground truths for the linguistic units in the speech data. In contrast, real speech is, by definition, more natural, and comes with higher within- and across-speaker acoustic variability. This is especially due to crowd-sourced nature of the speech audio in the existing audio-visual datasets. Therefore it was also of interest whether analysis findings from synthetic data would generalize to real speech, but also whether the VGS models trained on synthetic data would generalize to real speech and vice versa, as this also affects the general applicability of synthetic data for computational research on language learning in general.

For synthetic training and evaluation data, we used SPEECH-COCO dataset (Havard et al., 2017) based on MSCOCO (Lin et al., 2014). MSCOCO was originally collected to train computer vision systems, and consists of images paired with their verbal descriptions provided by human subjects. The dataset focuses on object recognition in context, and thus provides a variety of images of scenes and objects commonly observed in everyday life. The dataset includes a total of 123,287 images and covers 11 super-categories (e.g. animal, food, furniture etc.) and 91 common object categories (e.g. dog, pizza, chair) of which 82 categories contain more than 5,000 labeled samples (cases). Each image is paired by at least five written captions describing the scene using the object categories. SPEECH-COCO (Havard et al., 2017) was derived from MSCOCO by using a speech synthesizer to create spoken captions for more than 600k of the image descriptions in the original MSCOCO dataset (Chen et al., 2015). The speech was generated using a commercial Voxygen text-to-speech (TTS) system, which is a concatenative TTS system with four UK and four US English voices. SPEECH-COCO has the same datasplit as in MSCOCO 2014; the training set includes 82,783 images with corresponding 414,113 image descriptions and the validation set consists of 40,504 images paired with 202,654 captions. Each audio sample comes with synthesizer metadata on the audio caption, such as timestamps and identities of phones, syllables, and words synthesized, and we treated these as the gold standard phonetic reference of our speech data.

In our experiments, we randomly sampled two sets of 5k images from the original SPEECH-COCO validation set to be used for model validation and test data. The rest of the validation set (~30k images) were included in the training data. As a result, there were a total number of 113,287 images and 566,432 spoken captions for training, and two sets of 5,000 images with 25,000 utterances for validation and testing. In the linguistic representation analyses, one randomly chosen caption was used for each test image.

For real speech -based model training, we used Places400K corpus. It is based on Places205 image database (Zhou et al., 2014) that contains over 2.5 million images illustrating 205 different everyday scene types. Places Audio Caption (English) 400K data (Harwath et al., 2016) contains approximately 400,000 speech captions created for an equal number of images from Places205. Audio captions were collected from hundreds of speakers through Amazon's Mechanical Turk. During the data collection process, the user was asked to provide a free-form speech for each image describing the salient objects in it. There is only one verbal description per image, but compared to SPEECH-COCO, the average duration of the utterances is longer. The authors of Places400K automatically transcribed the spoken captions using automatic speech recognition (ASR) and reported approximately 23% word error rate for the results. Since we only use the text captions for the semantic retrieval analysis, and since there is no obvious reason why the ASR errors would bias the relative comparison of alternative models in the task, we find quality of the captions acceptable for the purpose. We split 10,000 validation and 10,000 testing images-caption pairs from the full dataset, and used the rest (392,385 image-speech pairs) for model training. In contrast to SPEECH-COCO that only consists of 8 different synthetic voices, Places400K represents notable variety in speakers and speaking styles due to its crowdsourced nature. Hence, their use in our experiments allows us to probe the impact of acoustical variety on learned model representations.

Since Places400K does not have existing phonetic annotations, we used a third corpus to investigate model representations with real speech. For this purpose, the so-called "*Large Brent*" subset (see Rytting et al., 2010) of Brent-Siskind corpus (Brend & Siskind, 2001) was used. The corpus consists of recordings of infant-caregiver interactions from four preverbal babies. The transcripts of the adult speech were transformed into phone- and word-level annotations using ASR-based forced-alignment by Rytting et al. (2010). In Räsänen et al. (2018), the transcripts were further syllabified based on the phone strings, and we use the 6,253 utterances with phone-, syllable- and word-level annotations as described in that paper. In contrast to SPEECH-COCO and Places400K, audio quality of Brent is significantly worse due to its at-home recordings. It also represents very different speaking style from the two other corpora. Therefore it was of interest to compare analysis results from SPEECH-COCO synthetic speech to those of Brent. Note that since Brent consists of audio only, it was not possible to evaluate audiovisual retrieval on that corpus.

The three datasets and their roles in model training, validation, audiovisual search evaluation, and model representation analysis are summarized in Tables 1 and 2.

**Table 1.** ***Datasets used for audiovisual model training (train), early stopping and model selection (dev), and analysis of audiovisual semantic retrieval (test). N refers to the number of utterances and corresponding spoken captions.***

| | *N* train | *N* dev | *N* test | speakers | speaking style |
|---|---|---|---|---|---|
| **Places400K** | 382,385 (images & utt.) | 10,000 | 10,000 | 2,683 | crowsourced real |
| **SPEECH-COCO** | 113,287 (images) x 5 (utt.) | 5,000 | 5,000 | 8 (4 x US, 4 x UK) | synthetic |

**Table 2.** ***Datasets used for analyzing model representations with respect to linguistic annotations. SPEECH-COCO test set is the same as the one used to test audiovisual semantic retrieval (listed in Table 1).***

| | | | | | phones | | syllables | | words | |
|---|---|---|---|---|---|---|---|---|---|---|
| | *N* test | duration | speakers | style | *types* | *tokens* | *types* | *tokens* | *types* | *tokens* |
| **SPEECH-COCO** | 5,000 | 214 min | 8 (4 x US, 4 x UK) | synthetic | 47 | 190,629 | 232 | 51,855 | 168 | 37,558 |
| **Brent** | 6,253 | 93 min | 4 | real IDS | 44 | 71,569 | 113 | 13,849 | 86 | 13,218 |

### *Model Training*

Since all inputs to the models had to be of equal length, we first zero-padded/truncated input log-Mel spectra to the length of 1024 frames (10.24 s) and 512 frames (5.12 s) for the Places and COCO datasets, respectively. Following the literature, the embedding space dimensionality *D* was set to 1024 for Places (Harwatha & Glass, 2017) and 512 for COCO (Chrupała et al., 2017). The resulting model parameter counts were approximately 21.2M (CNN0), 9.8M (CNN1), and 10.1M (RNN) parameters on Places, and 13.9M (CNN0), 7.4M (CNN1), and 8.8M (RNN) on COCO.

For model training, we used mini-batch size of 120 triplets and shuffled mini-batch sample assignments after each epoch. A new set of negative samples was also drawn for each epoch. Adam optimizer with a fixed learning rate of to 1e-4 was used. Early stopping based on development set recall@10 score with patience of 5 was used to control the training, and the best model according to the validation recall was then used for testing purposes. In all models, rectifier linear units were used as activation functions for all convolutional layers and hyperbolic tangents were applied for recurrent layers. Based on pilot experiments with several triplet loss margins, a margin *M* = 0.2 was ultimately chosen based on its superior audiovisual retrieval performance.

### *Evaluation Protocol*

For speech-to-image and image-to-speech retrieval tasks, we measured recall@10 for randomly sampled subsets of 1,000 image-caption pairs from our test data sets. For Places, we used all 10,000 test set image-caption pairs and sampled 1,000 pairs, measured the recall, and repeated the processes until distribution of recall scores for all testing samples converged to a normal distribution. We report recall scores then based on the mean and standard deviation of the obtained distribution. For COCO, we used all the possible 25,000 image-caption pairs (5,000 unique images, each paired with 5 captions), first divided them to five subsets of 5,000 image-caption pairs, and then sub-sampled a random set of 1,000 image-caption pairs for each subset until the mean recall@10 scores across all subsets converged.

For objective evaluation of audio-to-audio search results, semantic relatedness score (SRS) was calculated on the orthographic test set captions of both corpora. On Places, all the 10,000 test utterances were used, whereas one randomly chosen caption was used for each of the 5,000 images on COCO. COCO captions were textual to begin with, and we used the Places captions generated by Harwath et al. (2016) using ASR. For each of the test set query utterances, the corresponding textual caption was compared to the captions of the top 5 retrieved utterances using the SRS score in Eq. (3). As a reference, the process was repeated for 5-top dissimilar and 5-random captions. The analyses were conducted for all test set captions whose content words passed spell checking based on the Word2Vec model. This left us with 9,242 and 4,877 utterances for Places and COCO test sets for semantic similarity measurement, respectively.

For linguistic analyses, audio encoder activations were first recorded for all utterances in the test corpora. Similarly to Alishahi et al. (2017), activations were averaged across the duration of each annotated phone, syllable, or word token, so that each resulting activation sample corresponded to one linguistic token at the given level of analysis. All unit types with less than 50 tokens in a test set were then discarded from the analyses, and the resulting type and token counts are summarized in Table 2. Classifier-based separability analyses were conducted for all the tokens of a test corpus, where 80% of the tokens were used for training and 20% for testing of the classifiers (ensuring that the tokens in training and testing were from different utterances). For the KNN, $k = 15$ nearest neighbors were used based on initial optimization on a subset of data. Node selectivity and clustering analyses were conducted for a random sample of 50 tokens from each type, sampling uniformly and randomly from the full pool of test set tokens. This was done to ensure that the reported metrics reflect equally all the phonetic/syllabic/lexical types instead of being strongly biased towards the most frequent ones. The same random samples were used for all models and layers. Note that the classifier-based UAR metric is inherently unaffected by test class frequencies, and no such sampling procedure was needed for the classifier analyses. For the temporal segmentation analyses, original activations for all utterances in the test corpora were used. In addition, utterance onsets and offsets were automatically scored as correctly detected (and not counting any additional algorithm boundaries at those locations as insertions), as their detection can be considered as trivial due to the definition of an utterance as a stretch of speech separated by pauses or a change in speaker turn.

As a reference point, we also report measures obtained from the same models before their training (i.e., using the initial random parameters). This allows us to disentangle any effects of audiovisual learning from the potential benefits of simply performing a series of random non-linear transformations on the input speech data (see Chrupała et al., 2020, for a discussion).

## Results

Results of the experiments are divided into two parts: first, we ensure that all three model variants have learned the audiovisual mapping problem, and investigate their relative performance in capturing the semantics between the two modalities. In the second part, we focus on the internal representations used by the models in the multimodal learning task.

### Validation of Model Performance

We first ensured that training of all models converged to a meaningful solution of the audiovisual learning task by examining their validation losses and recall@10 scores (Fig. 4). This was also the case, and all three models obtained quite comparable training and validation losses and recall scores on both corpora despite their architectural differences. The only exception to the rule was CNN1, which exhibited superior recall and slightly lower loss on Places validation set compared to the CNN0 and RNN models. Monotonic convergence of all the measures suggests that there was no overfitting in any of the models.

The corresponding test set recall@10 measures on both speech-to-image and image-to-speech search tasks are shown in Table 3. The results for all models are very close to each other, with exact ranking depending on the dataset and task type. In general, the performance between the present models and those reported by Harwath et al. (2016) and Harwath and Glass (2017) are also within similar range. However, exact comparison is not possible, as the details of the test set were not identical due to different sampling strategies. For an unknown reason, our implementation of CNN0 replicated from Harwath and Glass (2017) falls behind the original study on both speech-to-image and image-to-speech search on Places. On the other hand, our CNN1 is similar to results from Harwath and Glass (2017) in performance. RHN-RNN results by Chrupała et al. (2017) are also shown in Table 3 as a reference, although they used different synthesized captions for the COCO data and a larger image search space. In general, the within-corpus results show that the three compared models all succeed in the audiovisual learning task, and they do so with a comparable performance.

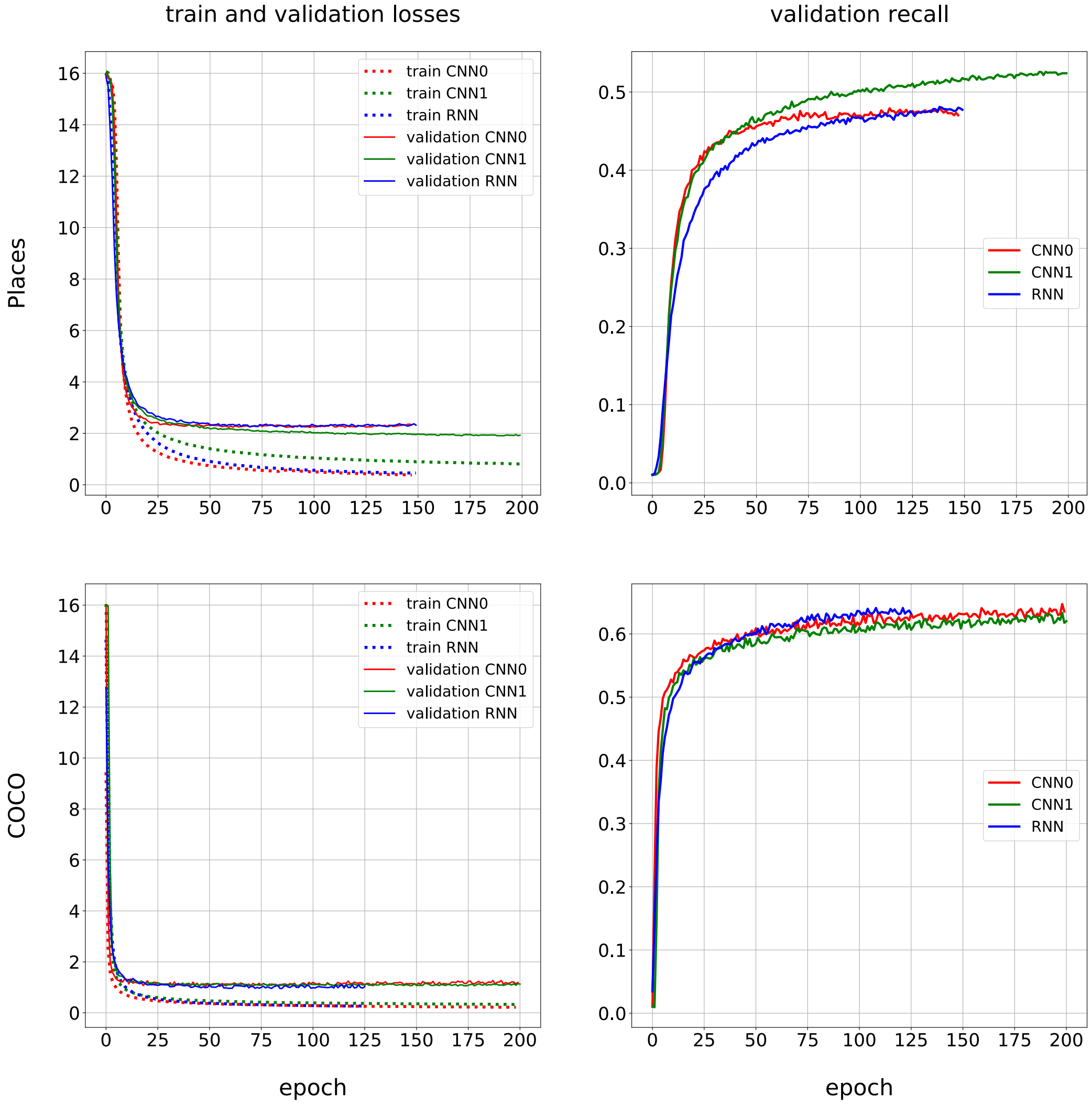

**Figure 4.** ***Training and validation losses and validation recall@10 scores for the three compared models as a function of training epoch. Top: Places corpus. Bottom: COCO corpus. Left: triplet loss scores. Right: recall@10 scores.***

**Table 3:** ***Recall@10 values obtained for "speech-to-image" (left) and "image-to-speech" (right). Means and SDs across different samplings of 1,000-sample search spaces are shown. Results from earlier two models tested on Places400K are shown as reference (*) (updated numbers from Harwath et al., 2018instead of the original papers). Note that the experimental setups are not identical, and therefore detailed comparison of numbers is not possible. Results from Chrupała et al. (2017) on another synthesized corpus, SS-COCO, are also shown (**), although they used a larger search space of 5,000 images in their experiments.***

| *<trainset>*-*<testset>* | speech-to-image | image-to-speech |
| --- | --- | --- |
| Places-Places | | |
| CNN0 | $0.473 \pm 0.015$ | $0.472 \pm 0.015$ |
| CNN1 | $0.522 \pm 0.016$ | $0.525 \pm 0.016$ |
| RNN | $0.466 \pm 0.015$ | $0.479 \pm 0.016$ |
| COCO-COCO | | |
| CNN0 | $0.643 \pm 0.015$ | $0.660 \pm 0.015$ |
| CNN1 | $0.633 \pm 0.015$ | $0.663 \pm 0.015$ |
| RNN | $0.643 \pm 0.014$ | $0.671 \pm 0.016$ |
| Places-COCO | | |
| CNN0 | $0.172 \pm 0.011$ | $0.180 \pm 0.012$ |
| CNN1 | $0.236 \pm 0.012$ | $0.234 \pm 0.013$ |
| RNN | $0.140 \pm 0.011$ | $0.175 \pm 0.011$ |
| *Harwath et al. (2016) (Places) | 0.548 | 0.463 |
| *Harwath and Glass (2017) (Places) | 0.564 | 0.542 |
| **Chrupała et al. (2017) (SS-COCO) | 0.444 | |

In terms of models trained on Places and tested on COCO (bottom part of Table 3, the performance of the models degrades substantially from the within-corpus experiments, even though the performance is still far above chance-level (0.01). Whether this is due to differences in acoustic signals (synthetic vs. real speech) or semantics of the images (common objects vs. scenes of "places") is unclear, although linguistic analyses discussed in section Results suggest that the acoustic mismatch may not be the issue. Among the models, CNN1 generalizes across corpora somewhat better than the CNN0 and RNN models.

## Qualitative Analysis of Semantic Retrieval

To further investigate how speech embeddings capture semantic similarities between image and speech modalities, we manually verified a number of retrieval results using the embeddings derived from the speech or image data. Table 4 shows an example of speech-to-speech search obtained using the speech embeddings. As can be observed, the first five most similar captions are semantically connected to the query caption. In the extracted examples, the query utterances and resulted utterances either include same objects or activities or share the same super-category (food, animal, etc.). Note that this matching is not possible by trivial matching of acoustic patterns alone due to lack of temporal alignment between the utterances. This indicates that the model has learned to link semantically similar utterances to each other without any supervised learning.

**Table 4.** ***Example output for speech-to-speech search using the CNN1 model on Places corpus. Query utterance and utterance transcripts corresponding to the five closest utterance embeddings are shown (spelling as they appear in the ASR-generated transcriptions).***

| query | **two cars traveling on a narrow suspension bridge under a blue cloudy sky.** |
|---|---|
| 1 | the sky is blue with lots of clouds. |
| 2 | the parking lot is empty the sky is cloudy. |
| 3 | a blue fairy goes under a blue bridge in a body of water on a clear sunny day with white clouds in the background and green tree surrounding it. |
| 4 | train crossing large bridge over bright blue water by the sun's going down with orange blue cloudy skies. |
| 5 | a blue cloudy sky bass casting a dark shadow on a roll of cars that are on the street and heavy traffic and there are trees on the side of the road. |

**Table 5.** ***Example output for speech-to-image search using the CNN1 model on Places corpus. Caption of the query utterance (as it appears in annotation files) and images corresponding to the five closest image embeddings are shown.***

**a small girl with blue and white striped shirt play.**

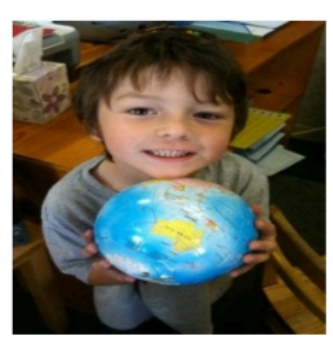
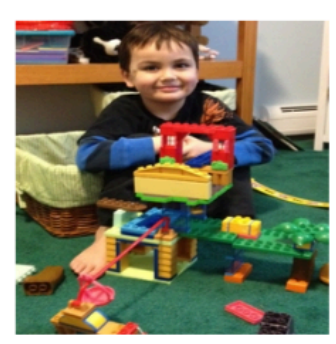
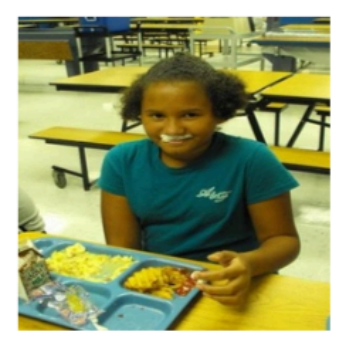
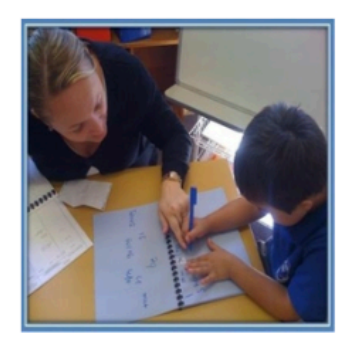
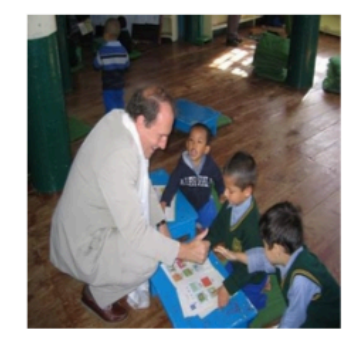

**Table 6.** ***Example output for image-to-speech search using CNN1 model on Places corpus. Left: query image. Right: transcripts of the utterances corresponding to the five closest utterance embeddings (spellings as they appear in annotation files).***

| query image | 5 highest scoring audio captions |
|---|---|
| 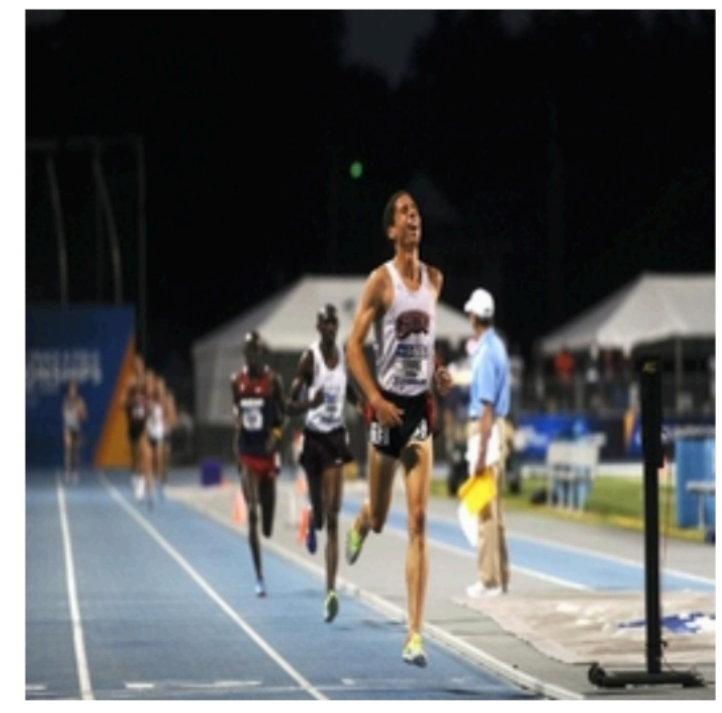 | • google people are racing around checked the first mermaid is wearing a blue uniform.<br>• pictures of man running in a race on the right track if people are around and starts with hurdles on the right.<br>• a woman running on a track there are people looking on in the center of the track are cameron and co.<br>• and there are several people standing along the back of the truck watching them.<br>• two women are running around a track there one is wearing blue and red and the other is black and white. |

Table 5 illustrates an example when spoken captions are used to find five best matching images. In the majority of observed samples of speech-to-image search, the resulting pictures contain objects corresponding to one or more of the content words spoken in the query caption. Finally, Table 6 shows five top similar captions resulting from a search based on a query image. In this example, as in the most cases of image-to-speech search results, the extracted captions are semantically related to the query utterance. While these examples are shown for the CNN1 model only, we manually verified that all the three models were able to extract semantic relations between audio captions and images in a qualitatively similar manner.

### Evaluation of Semantic Relatedness

By using our Word2Vec and SBERT-based measures of semantic relatedness, we calculated semantic similarity scores between captions corresponding to the five closest, five furthest, and five random embeddings with respect to every possible query utterance and corresponding caption drawn from the test set. For SRS, the measurement was done separately for all content words, and for content words after removing orthographically matching words from the compared captions. Table 7 shows the mean and SD SRS and SBERT similarity scores from the analysis. Fig. 5 also illustrates the obtained distributions of semantic similarities, including SRS with and without the repeating content words, when using the CNN1 model (results for the CNN0 and RNN are essentially similar and hence not shown separately). To further measure the degree that each model managed to capture the semantics of the data, we also quantified the difference in SRS and SBERT similarity score distributions between the top 5 similar captions and the 5 most distant/random captions using the Wilcoxon ranksum statistic. The corresponding test statistics are reported in Appendix C.

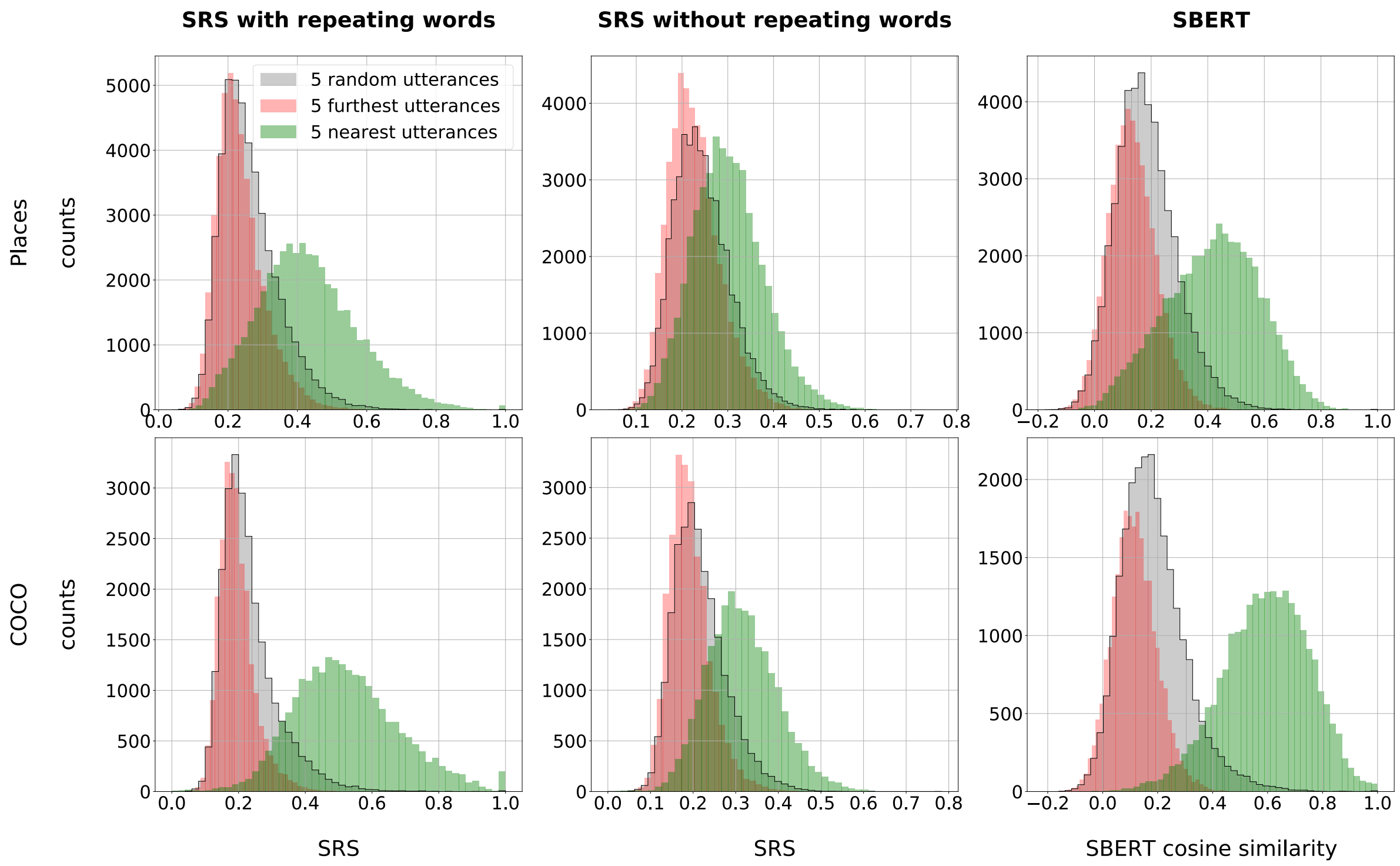

**Figure 5.** ***Semantic relatedness scores (SRS) and SBERT similarity scores for speech-to-speech embedding search results for the CNN1 model. The graphs show the distributions of similarities between query utterances and the five nearest, the five most distant, and five random captions collected for all test utterances.***

In all the models, semantic similarity of the nearest utterances is significantly higher than in a random sample of utterance pairs ($p < 0.001$ for all comparisons; see Appendix C for details). This is also the case after ignoring repeated words in the semantic similarity calculation. On the other hand, semantic similarity between the query caption and captions of the most distant embeddings are largely overlapping with distances to random embeddings. This likely reflects the use of margin in the VGS model loss function in Eq. (2), which basically constrains the model to focus on the structure of the multimodal embedding space only in the neighborhood of each data point. In contrast, different "degrees of semantic unrelatedness" are not captured by the model, as long as the embeddings of unrelated input pairs are already sufficiently distinct. Although this feature is already implicitly built in to the loss function, the SRS and SBERT metrics quantitatively demonstrate that the effect also seems to take place in practice.

As for the difference between SRS and SBERT, the SBERT model produces higher average similarity score for the closest utterances and lower average score for the furthest and random utterances compared to SRS model, reflecting its higher capacity in capturing semantics of full sentences with sentence-level embeddings. However, overall there is clear qualitative resemblance between SBERT and the SRS scores with repeating words (see Table 7).

**Table 7.** ***Medians (Mdn) and standard deviations (SD) of SRS and SBERT scores in case nearest, furthest, and random embeddings w.r.t. query utterances. Left: SRS using all content words in the utterances. Middle: SRS for content words excluding repeating words between query and search result utterances. Right: SBERT scores for full captions.***

| | **SRS with repeating words** | | | | | | **SRS without repeating words** | | | | | | **SBERT** | | | | | |
|---|---|---|---|---|---|---|---|---|---|---|---|---|---|---|---|---|---|---|
| | 5 nearest | | 5 furthest | | random | | 5 nearest | | 5 furthest | | random | | 5 nearest | | 5 furthest | | random | |
| | Mdn | SD | Mdn | SD | Mdn | SD | Mdn | SD | Mdn | SD | Mdn | SD | Mdn | SD | Mdn | SD | Mdn | SD |
| **COCO** | | | | | | | | | | | | | | | | | | |
| **CNN0** | 0.49 | 0.16 | 0.2 | 0.06 | 0.21 | 0.09 | 0.31 | 0.08 | 0.2 | 0.05 | 0.21 | 0.06 | 0.59 | 0.16 | 0.14 | 0.09 | 0.17 | 0.12 |
| **CNN1** | 0.51 | 0.16 | 0.19 | 0.05 | 0.21 | 0.09 | 0.31 | 0.08 | 0.19 | 0.05 | 0.21 | 0.06 | 0.6 | 0.16 | 0.11 | 0.08 | 0.17 | 0.12 |
| **RNN** | 0.49 | 0.16 | 0.21 | 0.06 | 0.21 | 0.09 | 0.31 | 0.08 | 0.2 | 0.05 | 0.21 | 0.06 | 0.59 | 0.16 | 0.15 | 0.1 | 0.17 | 0.12 |
| **Places** | | | | | | | | | | | | | | | | | | |
| **CNN0** | 0.37 | 0.13 | 0.23 | 0.07 | 0.25 | 0.09 | 0.29 | 0.07 | 0.22 | 0.06 | 0.24 | 0.06 | 0.4 | 0.17 | 0.14 | 0.09 | 0.17 | 0.11 |
| **CNN1** | 0.38 | 0.13 | 0.22 | 0.07 | 0.25 | 0.09 | 0.29 | 0.08 | 0.22 | 0.06 | 0.24 | 0.06 | 0.43 | 0.16 | 0.13 | 0.08 | 0.17 | 0.11 |
| **RNN** | 0.36 | 0.13 | 0.23 | 0.07 | 0.25 | 0.09 | 0.29 | 0.07 | 0.23 | 0.06 | 0.24 | 0.06 | 0.4 | 0.16 | 0.15 | 0.09 | 0.17 | 0.11 |

### *Discussion on Semantic Retrieval Experiments*

Overall, the retrieval performance in terms of recall@10, the SRS scores, and the qualitative analyses together confirm that all three models had acquired basic understanding of the semantic relationships between continuous speech and the related visual images. In addition, the three models did so in a comparable manner in terms of our analysis metrics despite the architectural differences between the models. Moreover, the analyses with SRS while excluding repeating words indicates that the semantic similarities among spoken utterances were not merely driven by shared words between the utterances. Instead, the models had learned something about semantic relationships of different words through their occurrences in similar visual contexts. This provides a solid starting point for investigating how the models actually learned to represent the spoken language input as a part of their solution to the audiovisual learning task.

## Results from Linguistic Selectivity Analyses

Our primary research question was whether the VGS models exhibit signs of emergent linguistic organization. For this, we studied patterns of selectivity across different linguistic units in the hidden layers of our trained speech encoder models.

Fig. 6 shows the analysis results for models trained and tested on the synthetic COCO data, whereas Fig. 7 shows the same analyses for models trained and tested on real speech (Places and Brent corpora). In addition, Fig. 8 shows the results for the Places-trained model tested on COCO, as the result allows us to disentangle the effects of training data from test data characteristics. Layer numbers of each model correspond to the numbers denoted in Fig. 2. L0 stands for the input log-Mel features.

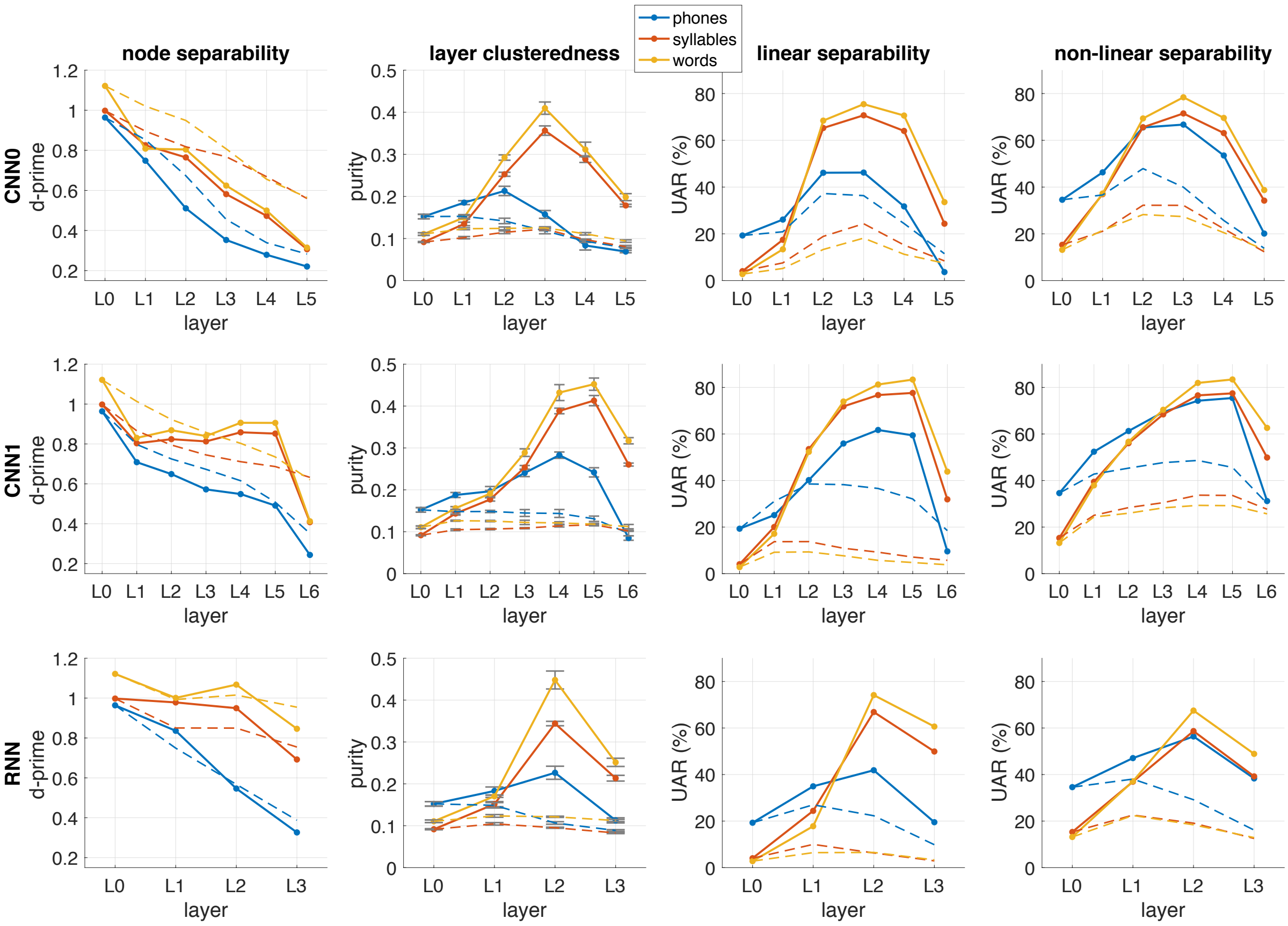

**Figure 6.** ***Analysis results for models trained and tested on COCO corpus. Each panel row corresponds to one of the models, CNN0, CNN1 or RNN, whereas columns correspond to the four studied selectivity metrics. Blue lines stand for phones, red for syllables, and yellow for words. Solid lines correspond to trained models and dashed lines for the corresponding baseline models before the training. Error bars for clusteredness represent SDs across different runs of the k-means analysis.***

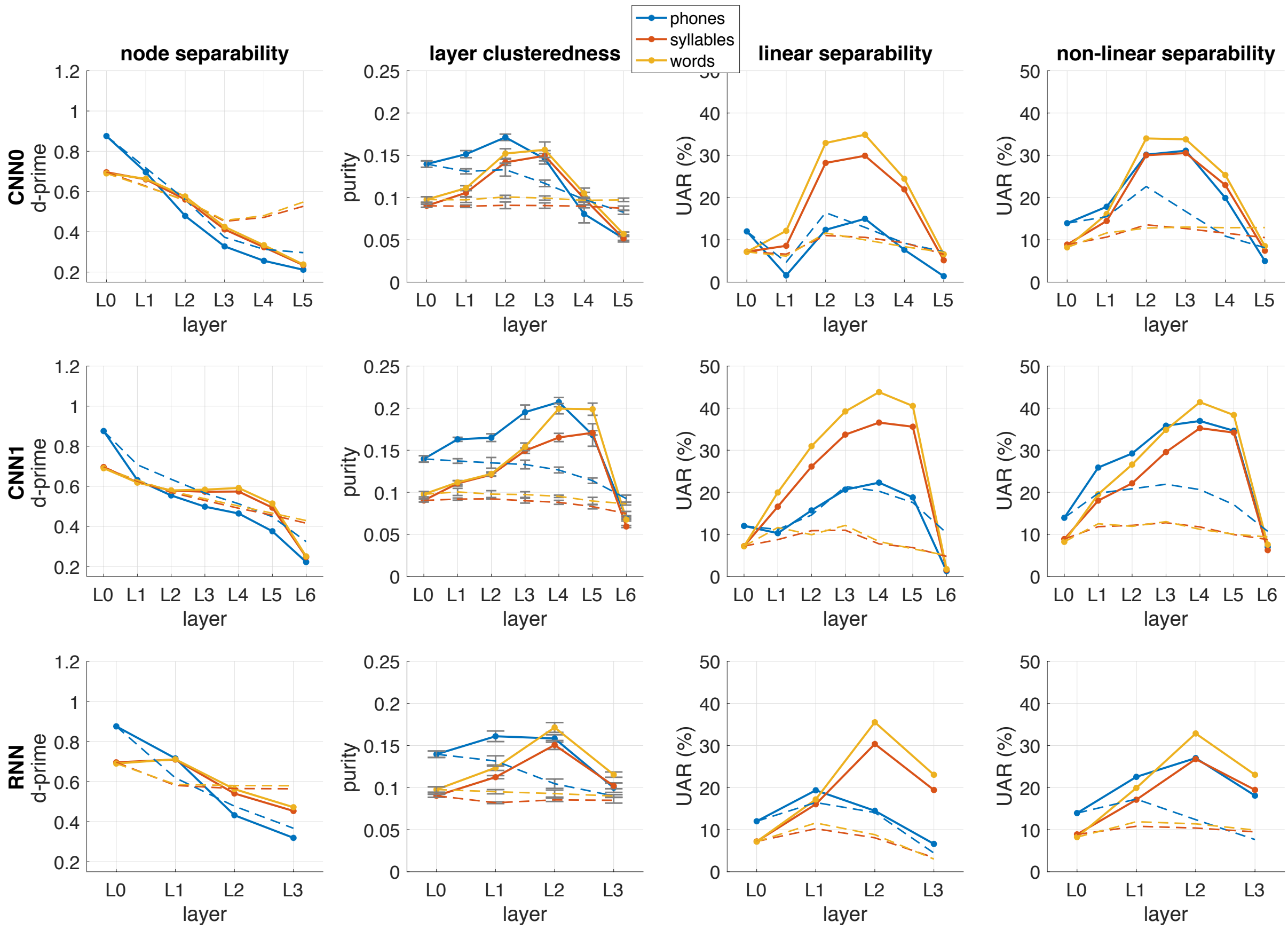

**Figure 7.** ***Analysis results for models trained on Places corpus and tested on Brent corpus.***

The first observation from the results is that the overall pattern of the analyses is very similar across the different combinations of training and testing corpora. Although the separabilities and classification accuracies are generally lower for Brent than for the synthetic speech from COCO, the relative measures from different layers of each model and in comparison to the untrained baselines are generally similar. Due to this, we will focus on discussing the general findings that hold across the different corpora, and separately mention whenever a finding only applies to a subset of training and testing conditions.

There are several patterns in the results that seem to be robust across the models and datasets. First of all, activation patterns of individual network nodes are poor at separating phone, syllable, or word types from each other, and this is true for all the three model variants (left columns in the figures). In CNN1 and RNN, there is a slight tendency of the node separability to improve in the early or middle layers due to model training. However, even in these cases, node separability is at its maximum or close to the maximum in the input layer. This means that the individual frequency channels of the Mel-frequency features are more informative of the underlying phonetic/syllabic/lexical identities than the individual nodes in hidden layers of the given models.

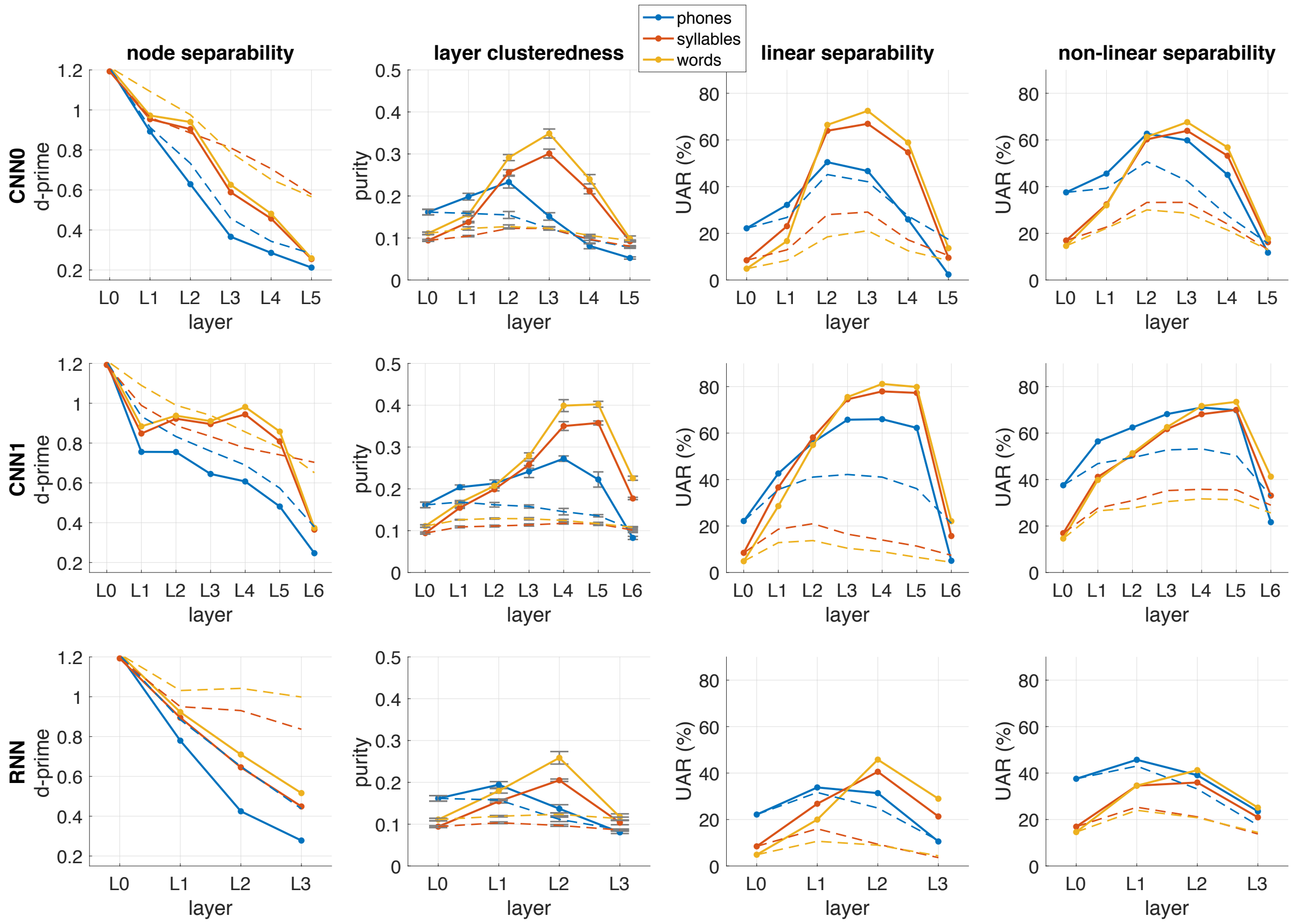

**Figure 8.** ***Analysis results for models trained on Places corpus and tested on COCO corpus.***

The picture changes when all the nodes of a layer are analyzed together (three last columns in each figure). Clusteredness and classification-based measures show clear effects of phonemic, syllabic, and lexical learning taking place in all the models. For instance, unsupervised clustering of CNN1 L4 activations achieves purity of nearly 0.3 in terms of phonetic categories on COCO data (Figs. 6 and 8), whereas clustering of the original log-Mel features leads to purity of 0.16 only. The effect is even larger at the syllabic and lexical levels. CNN1 L5 reaches lexical purity of approx. 0.45 with the vocabulary of 168 unique word types. Purities of RNN and CNN0 have similar patterns to those of CNN1, even though the exact layer in which the purity peaks differs from encoder architecture to others. Purities with the even more numerous syllables follow the same general trend, but with slightly lower purities. Notably, the improvements in syllabic and lexical clusteredness are not simply driven by the increasing receptive field sizes in deeper layers, as the there are no similar improvements in clustering purity for the representations extracted from the untrained but otherwise identical models. Purities measured on the Brent data are lower than those measured from the synthetic data, but there are still clear effects of training with a substantial improvement from the input features. We also conducted post-hoc analyses by measuring purities when the k-means clustering was initialized using the means of phone/syllable/word type activations, and the resulting purity scores were generally 0.1–0.3 higher than those observed for random initialization, but without changing the general pattern across different models, layers and datasets.

The linear and non-linear separability measures also indicate that there are large amounts of phonetic, syllabic, and lexical information encoded in the hidden layer activations of all three models. Representations derived from the middle layers (CNN0), middle to penultimate layers (CNN1), or penultimate layer (RNN) allow relatively accurate classification of syllables and words. On COCO training and testing, the non-linear classification reaches up to approx. 80% accuracy at all three levels of units with the CNN1 model activations, and the result is around 70% for an equivalent model trained on real speech. The corresponding accuracies for log-Mel features are below 40% for phones and below 20% for syllables and words. The qualitative pattern is similar to Brent test data, although the performance numbers are again lower due the more complex and noisy data.

The classification experiments also reveal that phonetic information seems to be embedded within the layers that also encode syllabic or lexical units (cf. the idea of overlapping representational planes in PRIMIR; Werker & Curtin, 2005). The maximum phone classification accuracy is often achieved for the same layers with the best performance on syllables and words, and not for the layers where the receptive field size best fits the typical phone durations (e.g., L2 and L3 in CNN1; see Fig. 2). In addition, phone classification is always clearly more accurate with the non-linear than linear classifier, whereas only minor differences between linear and non-linear classification are observed for syllables or words. This shows that phonetic information becomes the most refined in the same layers that encode syllabic and lexical information (i.e., is concurrently represented with higher levels of linguistic organization), suggesting that phonetic units become encoded in a context-sensitive manner. However, accurate decoding of the phone identities requires non-linear decoding of the activation patterns. Classification analyses also reveal that simply performing a number of random high-dimensional non-linear projections on the data seems to improve classification performance, as observed for the untrained models. However, the improvements are far from the benefits of audiovisual learning.

As for model comparisons, there are certain details that differ between the three architectures, even though all three architectures had very similar performance in the semantic retrieval tasks that they were trained for (see the previous sub-sections). In terms of CNN0 and CNN1, one difference is that phonetic clustering purity is higher for CNN1 than CNN0. In both models, the purity peaks at layers with comparable receptive field lengths (L2 in CNN0: 135 ms, L4 in CNN1: 165 ms) that are somewhat beyond typical phone durations. However, this may be explained by the higher number of nodes and thereby higher representational capability in CNN1 at the given layer (same 512 in all layers for CNN1), whereas CNN0 architecture uses increasing number of nodes as the receptive field increases (as specified in Harwath & Glass, 2017). In addition, non-linear separability of phonetic units is somewhat higher with CNN1 representations than those in CNN0. Patterns for words and syllables are more similar for the two models. In terms of comparison between the CNNs and the RNN, the results are remarkably similar despite their architectural differences. The first recurrent layer (L2) of the RNN is similar to the middle layers of the CNN models, and a similar drop in linguistic selectivity is observed for the last layer of RNN as in those of both CNNs.

**Results from Temporal Segmentation Analyses**

For the temporal segmentation analyses, we first compared L2-norm, entropy, and linear regression-based representations in the task and found that the regression approach led to somewhat higher segmentation performance than the other two. Due to this, we focus on the regression results here and the full set of L2-norm and entropy-based measures can be found from Appendix B.

Fig. 9 shows the temporal segmentation analysis results for models trained and tested on COCO. Figs. 10 and 11 show the corresponding results for models trained on Places and tested on Brent, and for models trained on Places and tested on COCO, respectively. Baseline performance levels with untrained models are also shown for reference.

Looking at the COCO-COCO results in Fig. 9, there are two key findings: 1) Segmentation performance of all types of units is far above zero, and CNN hidden layers have higher segmentation accuracy for phones and syllables than when using the log-Mel features. 2) Untrained model performance is also relatively high throughout the conditions. This shows that much of the temporal dynamics exhibited by the models (as captured with the present methodology) are already captured by the interaction of input features and non-linear processing steps. In other words, there is only a small effect of training on how the activation patterns reflect linguistic unit boundaries in time. On Places-Brent, the general pattern of results is again very similar to COCO-COCO, but in this case the metrics of the trained models are even closer to the same models with random initial parameters. This is not due to training with real speech, as the performance in the Places-COCO condition again reflects the pattern observed in COCO-COCO. In addition, overall scores for syllables and words are somewhat higher on Brent than on COCO test data. However, this is primarily explained by the substantially shorter average utterance length on Brent, which means that the relative proportion of trivial utterance onset and offset boundaries is much higher on Brent than on COCO.

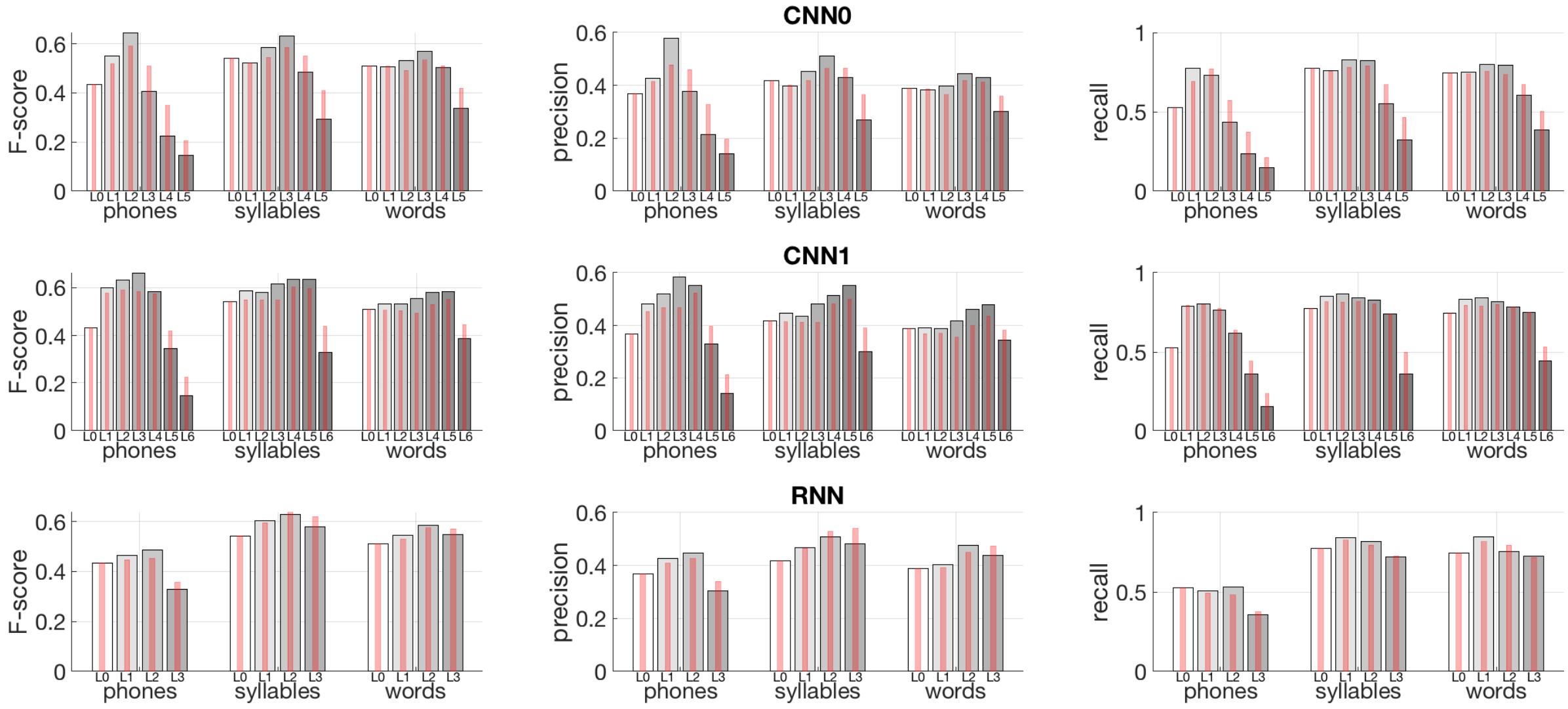

**Figure 9.** ***Segmentation results from models trained and tested on COCO corpus when using linear regression from activation magnitudes to unit boundary scores as the signal representation. Bars in each subplot represent layer-specific performance metrics for segmenting phones, syllables, and words, respectively. Different layers from input (L0) to last network layer are shown with different shade bars from left to right. Results are shown for F-score (left panels), precision (middle panels), and recall (right panels), and for the three tested models: CNN0 (top), CNN1 (middle), and RNN (bottom). Thin red lines denote baseline performance with untrained models.***

In terms of different layers, there is a small trend of earlier CNN layers to better reflect phonetic unit boundaries while syllable and word boundaries are better accessible from deeper layers. In addition, on COCO test data, the CNN models lead to higher phone and syllable segmentation performance compared to words, whereas on Brent syllables and words are actually more accurately segmented than phones. The best phone segmentation F-score of 0.66 is obtained by CNN1, followed by 0.65 for CNN0, both on COCO training and testing. The corresponding numbers for syllables are 0.64 and 0.63, respectively. The RNN has lower phone segmentation performance than the CNNs, whereas its syllable and word segmentation scores are generally comparable to those of other models (e.g., F-score of 0.64 for syllables on COCO-COCO).

Comparing the current regression-based segmentation results to the L2-norm and entropy-based representations reported in Appendix B, the results are largely qualitatively similar across the approaches. The clearest difference is a modest performance gain in syllable and word segmentation for the regression approach, as compared to the other two methods.

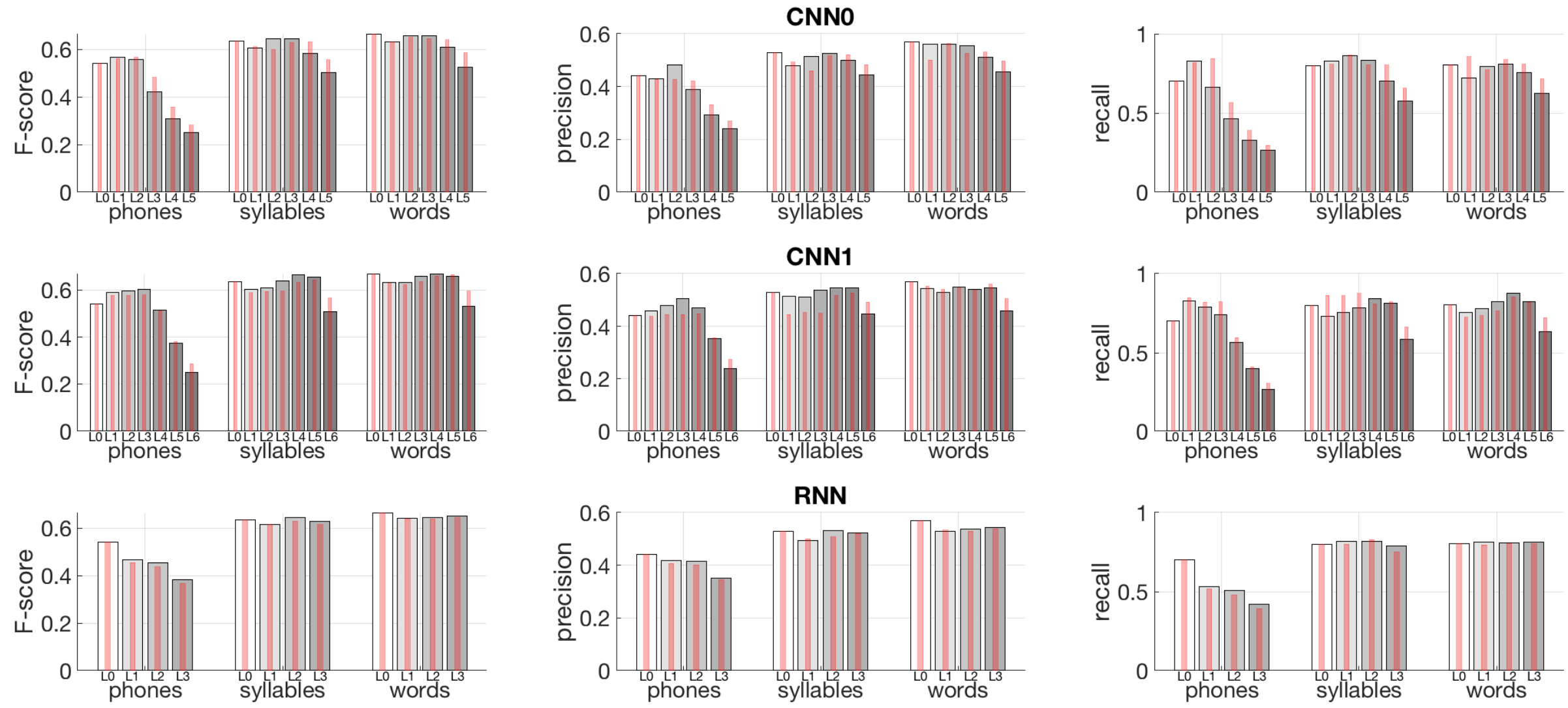

**Figure 10.** ***Segmentation results for models trained on Places and tested on Brent data.***

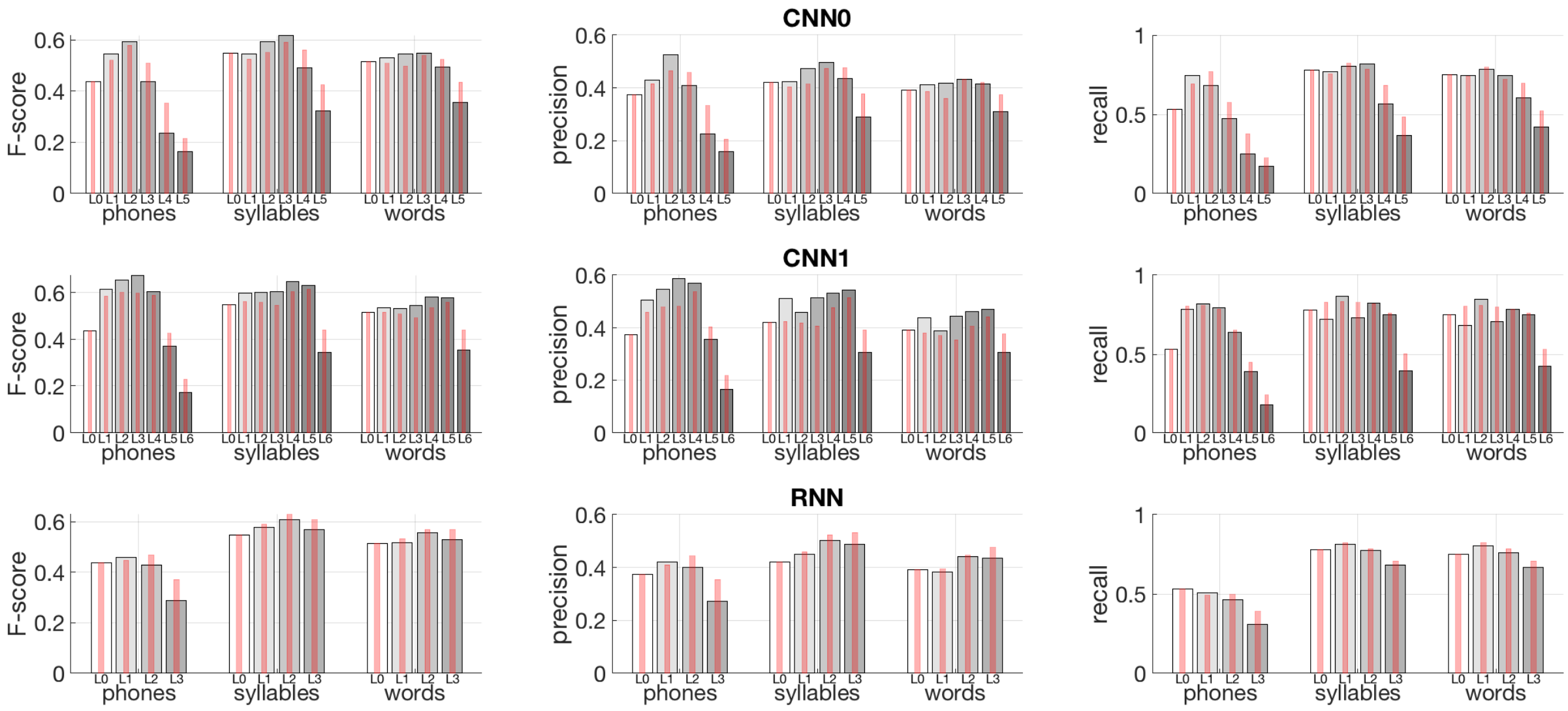

**Figure 11.** ***Segmentation results for models trained on Places and tested on COCO data.***

Compared to earlier studies, the observed segmentation scores are far from the F-scores of 0.72–0.80 reported for modern unsupervised phone segmentation algorithms (see, e.g., Hoang & Wang, 2015, for an overview), typically tested on TIMIT corpus (Garofolo et al., 1993). In addition, relatively low precision of the segmentation scores means that there is a substantial amount of oversegmentation involved in the process. In other words, the network dynamics have some correspondence to linguistic unit boundaries in time, but the segmentation behavior is far from being perfect. The observed performance is also somewhat worse than that reported by Harwath and Glass (2019) despite using the same model architecture and training protocol (CNN0). The reason for this difference is unclear, but may have to do with the different selection of the test data, as Harwath and Glass used read speech from TIMIT to test their model trained on Places. In general, it appears that temporal dynamics of the models are informative of phonetic, and to a smaller degree, syllabic and word boundaries. However, the effects of training are much less pronounced than with the selectivity analyses. Again, the main pattern of results does not seem to depend on whether the data are real or synthetic speech in nature.

## Discussion

This work set out to investigate whether cross-modal and cross-situational learning can give rise to emergent latent linguistic structure, as predicted by LLH. After formulating the LLH and reviewing the findings from the existing literature, we investigated the idea systematically using machine learning models relying on statistical dependencies between images and their spoken descriptions. This can be viewed as simulation of cross-situational learning (e.g. Smith & Yu, 2008) with a high degree of referential ambiguity, as the models had to automatically discover which aspects of the utterances (in time and frequency) were related to what kind of visual referents (in terms of visual features and spatial positions). We compared three distinct speech encoder networks in the task to understand how speech encoder architecture impacts the audiovisual mapping performance and the manner that the networks encode linguistic information. In addition, we compared learning from both synthetic and real speech. As we analyzed the models, we observed that all networks exhibited similar capabilities in learning the semantic structure between spoken language and image data. In addition, all networks showed clear signs of linguistic organization in terms of all three unit types of analysis, namely phones, syllables, and words.

In terms of our more detailed linguistic analyses, the present findings largely align with the earlier literature on investigating linguistic units in VGS models (e.g., Chrupała et al., 2017; Alishahi et al., 2017; Havard et al., 2019a, 2019b; Merkx et al., 2019). However, the present study is the first one to show that broadly similar learning takes place in different model architectures (convolutional and recurrent) and on both synthetic and real speech. In addition, for the first time, we analyzed the emergence of phonetic, syllabic and lexical representations with a shared set of metrics, and probing for both linear and non-linear separability of the representations in terms of underlying linguistic units in the input. The analysis revealed that all three levels of representation exist in the trained models in parallel. Also, while phone-level information is generally accessible already from the earlier layers, phonetic information also co-exists in the layers that also encode syllabic or lexical information, but the information requires non-linear decoding. In addition, the results clearly demonstrate that the linguistic information in these type of models becomes encoded as distributed representations, whereas informativeness of individual network nodes with respect to linguistic unit types is very limited. Temporal dynamics of the models also seem to carry information related to linguistic units in the input speech, especially phones and syllables, although it appears that only a limited proportion of this is actually driven by audiovisual learning in the models. However, this is not surprising, as many of the existing unsupervised phone and syllable segmentation algorithms can already perform relatively well by only analyzing the original signal-level acoustic changes (e.g., Hoang & Wang, 2005, and Räsänen et al., 2018, and references therein).

Comparing our selectivity results to the earlier work, Chrupała et al. (2017) and Merkx et al. (2019) found that some model layers specialized for lexical processing whereas some others encoded such information to a lesser degree. According to Chrupała et al. (2017), the accuracy for predicting presence of individual words in speech increases towards the deeper layers of the network but decreases slightly for the last layers of the model. Our results for layer selectivity in the RNN model generally replicate those findings, and similar behavior can be observed for the CNN models as well. Regarding phonemic representations, Alishahi et al. (2017) and Drexler and Glass (2017) found that phone-like information was most evident in the early-to-intermediate layers of their networks, and where deeper layers showed slightly decreasing phonemic specificity. In our results with RNN model (e.g., Figs. 6 and 7), phones are also the most prominent in the first recurrent layer or in the convolutional layer preceding the recurrent layers of our RNN model, when compared to the penultimate recurrent layer (L3 in our RNN model; cf. Fig 2). Moreover, as our results illustrate, a similar general pattern of initially increasing and then suddenly decreasing linguistic unit selectivity is observed for all the model architectures and for all linguistic units. While much of the earlier work has been carried out using RNN-based VGS models, our results also indicate that the findings also apply to CNN-based architectures with different temporal characteristics, demonstrating the robustness of the phenomenon.

In terms of data characteristics, the present study also shows that the selectivity analyses conducted on synthetic speech are qualitatively similar to those on real speech.

This also strengthens the findings from the earlier visual grounding studies that have used synthetic speech (e.g., Alishahi et al., 2017). While not all synthetic speech is equal in terms of naturalness and acoustic variability, the finding also suggests that modern high-quality speech synthesis could be used for speech data creation in other computational modeling studies when lacking suitable real-speech corpora. This is especially relevant for studies where input to the learner needs to be dynamically adjusted depending on the learner behavior, such as simulating learning in infant-caregiver interaction using computational agents (e.g., Asada, 2016) and references therein). However, the potential limitations of synthetic speech should still be carefully and separately considered for each study.

In summary, together with the earlier computational findings reviewed, the present study demonstrates that speech comprehension, as defined in terms of capacity to associate spoken language with its referential meaning, does not necessitate a priori parsing of the speech input into distinct units such as phones or words. Instead, a flexible statistical learning machine focusing on modeling the dependencies between different perceptual channels is sufficient for capturing rudimentary semantics of speech, at least in the present type of simplified audiovisual learning scenarios. In addition, when implemented as a neural network with several hidden layers, these hidden layers start to reflect selectivity towards different types of linguistic units that the input speech consist of. This is in line with earlier findings using neural network models using supervised training (Nagamine et al., 2015; see also Magnuson et al., 2020) or simplified visual input (Räsänen and Khorrami, 2019). Here we show that similar emergence of units can be observed in learning conditions analogous to cross-situational learning. This lends initial support for the LLH: the idea that infant language representation learning may be driven (or at least supported) by learning processes that do not directly aim at learning such representations, but where the linguistic representations become acquired as a byproduct of multimodal sensing and interaction with the environment.

Note that the idea of initially general (non-linguistic) perceptual processing and gradual phonological development enabled by concurrently developing lexicon is in line with PRIMIR theory (Werker & Curtin, 2005). However, PRIMIR also assumes that acoustic word-forms are segmented before being associated with their meanings. The present work together with the earlier reviewed studies (e.g., Alishahi et al., 2017; Chrupała et al., 2017; Havard et al., 2019a; Harwath and Glass, 2019; Merkx et al., 2019; Scholten et al., 2020) demonstrates that explicit segmentation into acoustic word-forms before linking them to their meanings is simply not needed. In contrast, both sub-lexical and lexical representations can gradually emerge from the interaction of rich multimodal experiences available to the learner, when the learner is simply optimizing the multimodal predictive value of the auditory input. This is also in line with computational models demonstrating that there are synergies in learning multiple levels of linguistic structure simultaneously (e.g., Feldman et al., 2013) or in word segmentation and meaning acquisition (Johnson et al., 2010).

We would like to emphasize that the present type of VGS models do not aim at modeling neurophysiology of speech perception, and exploring such a connection is greatly beyond the scope of the present study. In contrast, our present aim has been to investigate LLH at the level of computational principles. However, despite the conceptual gap between artificial and real neurons, ANNs have been successfully applied to modeling of cortical organization in case of visual (Yamins & DiCarlo, 2016) and auditory (Kell et al., 2018) processing. Therefore linking model data from VGS models to neurophysiological data could be attempted in the future, given access to suitable human data. On the other hand, it would be an interesting avenue to explore VGS model architectures further by taking into account known characteristics of the auditory and associative areas in the brain.

**Limitations of the Present Study**

From the point of view of human language learning, one of the main limitations of the present study is the data used in our experiments. While language learning children observe the world from their own visual perspective and predominantly hear spontaneous speech in interaction with their caregivers (e.g., Yurovsky et al., 2013), our model training datasets consisted of photographs and their verbal descriptions. This means that both the visual and auditory experiences differ from those of a child, and also the relationship between the two modalities is much more systematic than what would be expected from situated caregiver speech. While caregiver speech is not random with respect to otherwise observable environment, caregivers do not tend to narrate everything that the child is observing. On the other hand, factors such as joint attention, prosodic cues of child-directed speech (CDS), skewed statistics of the visual experiences (e.g., Clerkin et al., 2017), and gradual increase in the complexity of the speech input may help infants to resolve audiovisual referential ambiguity, whereas the present VGS models do not receive any "highlighting" of relevant targets in the visual or auditory domains. As for audio quality, the synthesized speech in COCO dataset necessarily has less acoustic variability than authentic caregiver speech, making the speech parsing problem easier. However, the experiments with Places corpus inevitably show that our primary findings also apply to learning from real speech, and similar patterns of linguistic unit emergence (albeit with smaller effects) can be seen when the models are tested on naturalistic IDS speech.

In terms of mere input quantity, our training set totaled to approximately 400 h

(COCO) or 895 h (Places) of descriptive speech paired with visual scenes. In comparison, an average infant hears approx. 3 hours of CDS per day (Bunce et al., in preparation). The total amount of speech heard by the first birthday would then correspond to approx. 1000 h of CDS, i.e., by time when the child starts to comprehend some tens of words (CDI data from Wordbank; Fenson et al., 2007; Frank et al., 2017). In this context, at least the approximate scale of magnitude between toddler language input and our model input does not seem totally implausible, assuming that some tens of percentages of CDS input would relate to situations with opportunities for visual grounding. In addition, the present models can also solve the audiovisual mapping problem with much less data with reasonable performance (we have also tested learning using only one caption per COCO image, corresponding to a total of 80 h of speech; not reported separately). However, for the reasons listed above (and many others, such as developmental factors), detailed comparison between infant input and our study is not feasible. It is also not meaningful, as the present hypothesis is not that language learning would *only* take place through audiovisual learning. In contrast, the main goal has been to investigate the degree that referentially-driven multimodal learning can, *in principle*, explain aspects of early language organization.

As another central limitation, our present data consisted of English speech only. While this necessarily limits the extent that conclusions can be drawn cross-linguistically, the use of English also limits the capability to disentangle syllable- and word-level representations from each other. This is since a substantial proportion of the English word tokens consists of monosyllabic words (Greenberg, 1999), and this was also the case for our present data.

Finally, our viewpoint to the structure learned by the networks is necessarily limited, even though we combined a number of measures in our experiments. For instance, we did not systematically compare the networks in behavioral experimental paradigms such as gating experiments (see Havard et al., 2019b), nor investigated the confusions among different linguistic unit types. In contrast, we focused on understanding the dynamics of the internal representations from the point of view of hierarchy of linguistic units of different granularity.

In the future work, it would be important to test the audiovisual models with real infant language and visual input. Some baby steps to this direction already exist (Räsänen & Khorrami, 2019), but systematic investigation at the scale of real infant language experiences would be ideal to understand the role of visual[4] experience in early organization of language. Ideally, datasets from several different languages would be also utilized and compared. In addition, comparison and combination of the VGS models with purely auditory predictive models (e.g., van den Oord et al., 2018) should be conducted to understand the relative roles of different perceptual modalities in early language learning.

[4] In the general case, this concerns parsing of speech in the context of sensory information from auditory, visual, somatosensory, and olfactory channels. In addition, motor activity and internal representations of emotions and interoception of basic bodily functions should be included as potential factors in speech grounding.

## Conclusions

One of the puzzles in the study of child language acquisition research is the question how infants learn to parse the noisy and highly variable acoustic speech input into a meaningful and structured interpretation of the spoken message, and then to gradually use the language in a compositional and generative manner. Several potential mechanisms have been proposed to solve problems such as phonemic categorization, word segmentation, and word meaning acquisition, but the overall picture of how these bits and pieces fit together remains unclear. It is especially unclear what would be the ecological or functional pressure for the baby brain to solve a series of proximal language processing sub-problems before being able to utilize the benefits of speech perception to comprehend the world around them.

The latent language hypothesis investigated in this paper is aimed to shed some light on this puzzle by proposing that linguistic knowledge could emerge from predictive optimization across sensorimotor modalities and across time. By doing so, separate solutions to the several intermediate speech parsing problems would not be necessary. The reviewed studies and the present findings demonstrate that the audiovisual aspect of LLH is, in principle, a potential mechanism for assisting in language representation learning. However, the link to behavioral data is currently limited, and therefore nothing conclusive can be said on human learning based on the models alone. Yet, the VGS models show that the role of multisensory input in the context of language learning should not be underestimated, and that access to rich multimodal experiences concurrent to speech input have the potential to assist the learners already in the first stages of language learning. In future work, it would therefore be important to better understand the role of multimodality but also purely auditory predictive processing in early language acquisition.

## Data, code and materials availability statement

### Program code and intermediate data files:

Python and MATLAB scripts for model definition, training, retrieval evaluation, and linguistic analyses, and result plotting are available at https://github.com/SPEECHCOG/VGS_XSL. Trained TensorFlow models, model hidden layer activations ("model outputs") used in the linguistic analyses, and the corresponding data annotations are available on Zenodo at https://doi.org/10.5281/zenodo.4564283.

### Datasets used in the study:

Datasets used in the study are publicly available for download from their original sources after registration to the respective sites:

Brent-Siskind corpus: https://childes.talkbank.org/access/Eng-NA/Brent.html
Places audio captions: https://groups.csail.mit.edu/sls/downloads/placesaudio/downloads.cgi
Places205 images: http://places.csail.mit.edu/downloadData.html
SPEECH-COCO audio captions: https://zenodo.org/record/4282267
MSCOCO images https://cocodataset.org/#download

The derived version of "Large-Brent" with utterance-level waveforms with their phone, syllable, and word-level transcripts (based on forced-alignment from Rytting et al., 2010, and syllabification in Räsänen et al., 2018) is available from the second author upon request. The data cannot be shared publicly, as it would require redistribution of modified (split + pruned) Brent-Siskind corpus audio files. Model outputs and the corresponding annotations for the Large-Brent data are available in the Zenodo-hosted datafile mentioned above.

## Authorship and Contributorship Statement

K.K. planned the study together with the second author. She was also responsible for model development and implemented all program code related to model training, retrieval evaluation, and initial linguistic selectivity analysis scripts. K.K. also prepared the data, ran all experiments except for the final linguistic analyses, and produced the corresponding result figures and tables. She wrote the first draft of the article together with the second author and participated to editing of the initial and revised manuscript versions.

O.R. was responsible for the initial idea of the study, for theoretical framing of the work, and generally supervising the work. He planned the work jointly with the first author and participated to methodological development. He also implemented the final linguistic analysis scripts (selectivity and temporal) and conducted and reported the corresponding experiments. O.R. wrote some sections of the first manuscript draft jointly with the first author and was largely responsible for editing the initial and revised versions of the manuscript.

## Acknowledgements

This research was funded by Academy of Finland projects no. 314602 and 320053. The authors would like to thank Herman Kamper for his help with VGS model training and for the highly useful discussions on the topic. The authors would also like to thank Grzegorz Chrupała for providing details on his earlier work.

Code used for model training and analysis, trained models, and hidden layer activation data are available for download at https://github.com/SPEECHCOG/VGS_XSL.

## Appendix A

Selectivity analyses (Fig. A.1) and temporal segmentation results (Fig. A.2) for models trained on COCO and tested on Brent corpus.

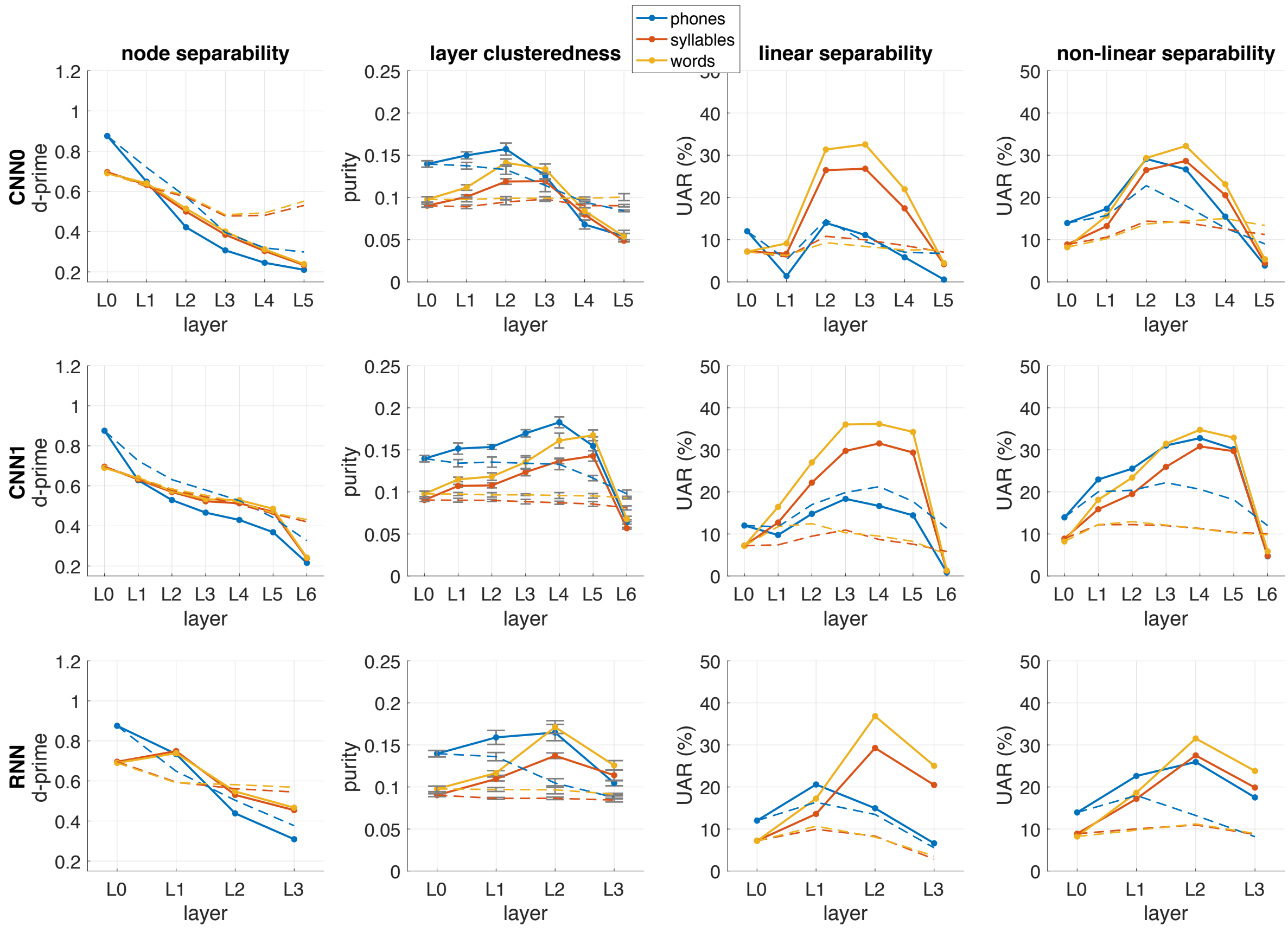

**Figure A.1.** ***Analysis results for models trained on COCO corpus and tested on Brent corpus. Each panel row corresponds to one of the models, CNN0, CNN1 or RNN, whereas columns correspond to the four studied selectivity metrics. Blue lines stand for phones, red for syllables, and yellow for words. Solid lines correspond to trained models and dashed lines for the corresponding baseline models before the training. Error bars for clusteredness represent SDs across different runs of the k-means analysis.***

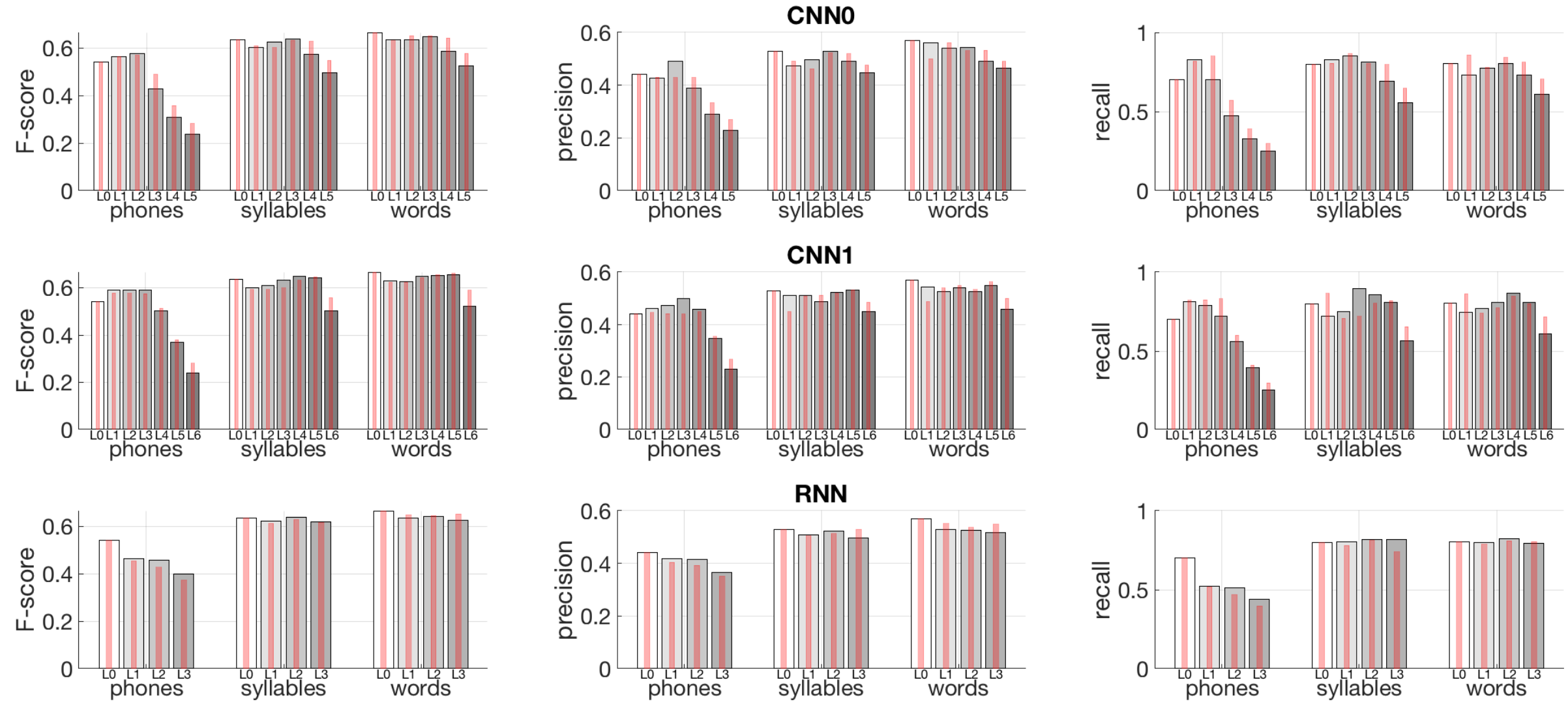

**Figure A.2.** ***Segmentation results from models trained COCO and tested on Brent corpus when using linear regression from activation magnitudes as the signal representation. Bars in each subplot represent layer-specific performance metrics for segmenting phones, syllables, and words, respectively. Different layers from input (L0) to last network layer are shown with different shade bars from left to right. Results are shown for F-score (left panels), precision (middle panels), and recall (right panels), and for the three tested models: CNN0 (top), CNN1 (middle), and RNN (bottom). Thin red lines denote baseline performance with untrained models.***

## Appendix B

Results for segmentation analyses using L2-norm or entropy of layer activations as the signal representation (instead of using the linear regression scores in the main results).

L2-norm measures: Fig. B.1: training and testing on COCO. Fig. B.2: training on Places and testing on Brent. Fig. B.3: training on Places and testing on COCO. Fig. B.4: training on COCO and testing on Brent.

Entropy-based measures: Fig. B.5: training and testing on COCO. Fig. B.6: training on Places and testing on Brent. Fig. B.7: training on Places and testing on COCO. Fig. B.8: training on COCO and testing on Brent.

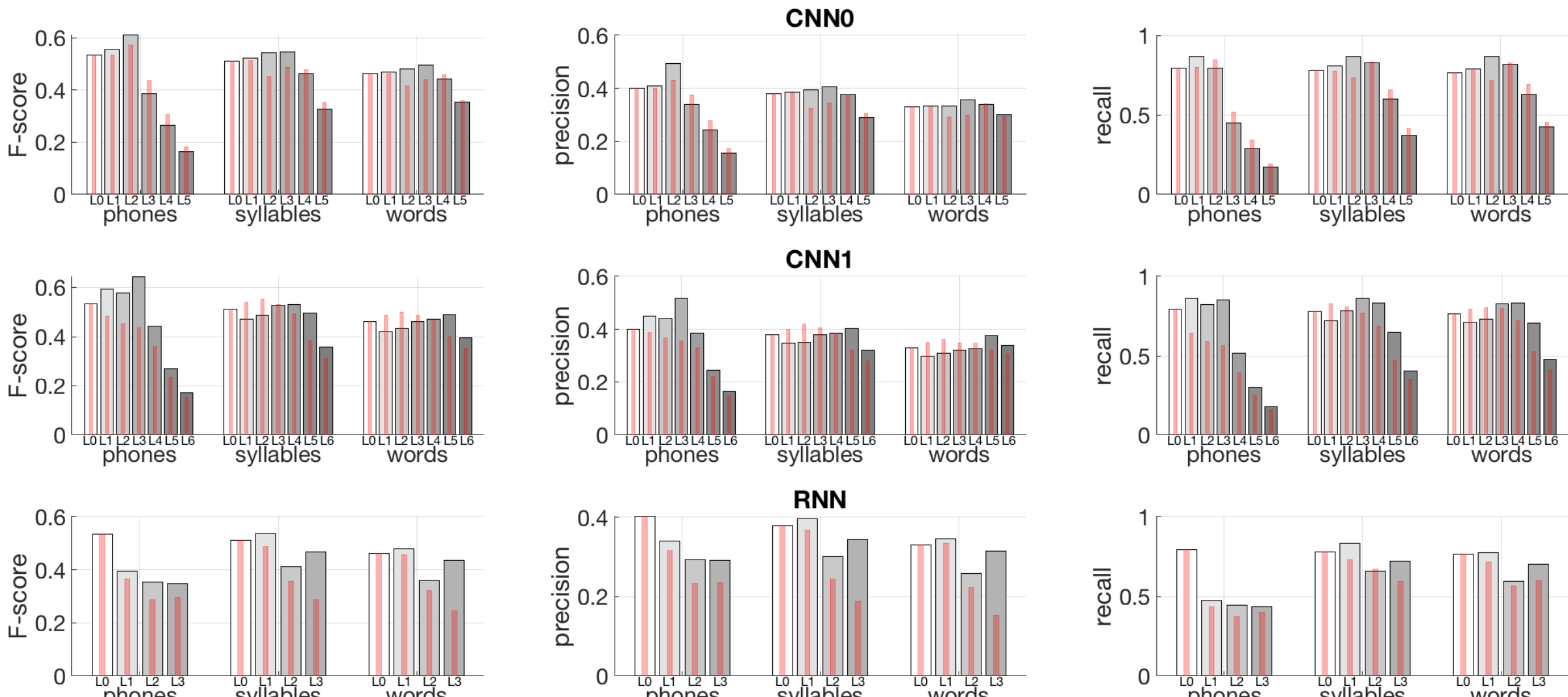

**Figure B.1.** ***Segmentation results from models trained and tested on COCO corpus when using L2-norm of activation magnitudes as the signal representation. Bars in each subplot represent layer-specific performance metrics for segmenting phones, syllables, and words, respectively. Different layers from input (L0) to last network layer are shown with different shade bars from left to right. Results are shown for F-score (left panels), precision (middle panels), and recall (right panels), and for the three tested models: CNN0 (top), CNN1 (middle), and RNN (bottom). Thin red lines denote baseline performance with untrained models.***

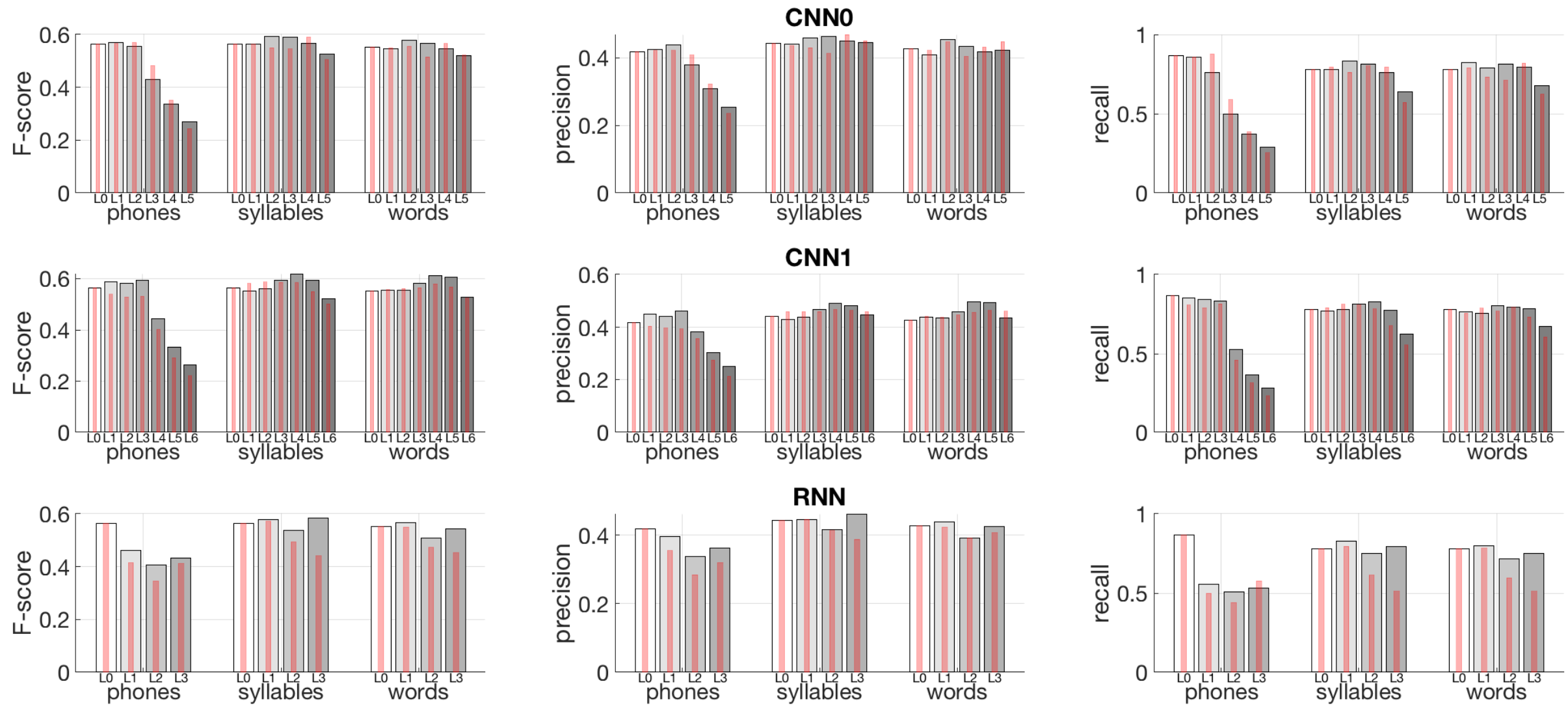

**Figure B.2.** ***Segmentation results for models trained on Places and tested on Brent data when using L2-norm of activation magnitudes as the signal representation.***

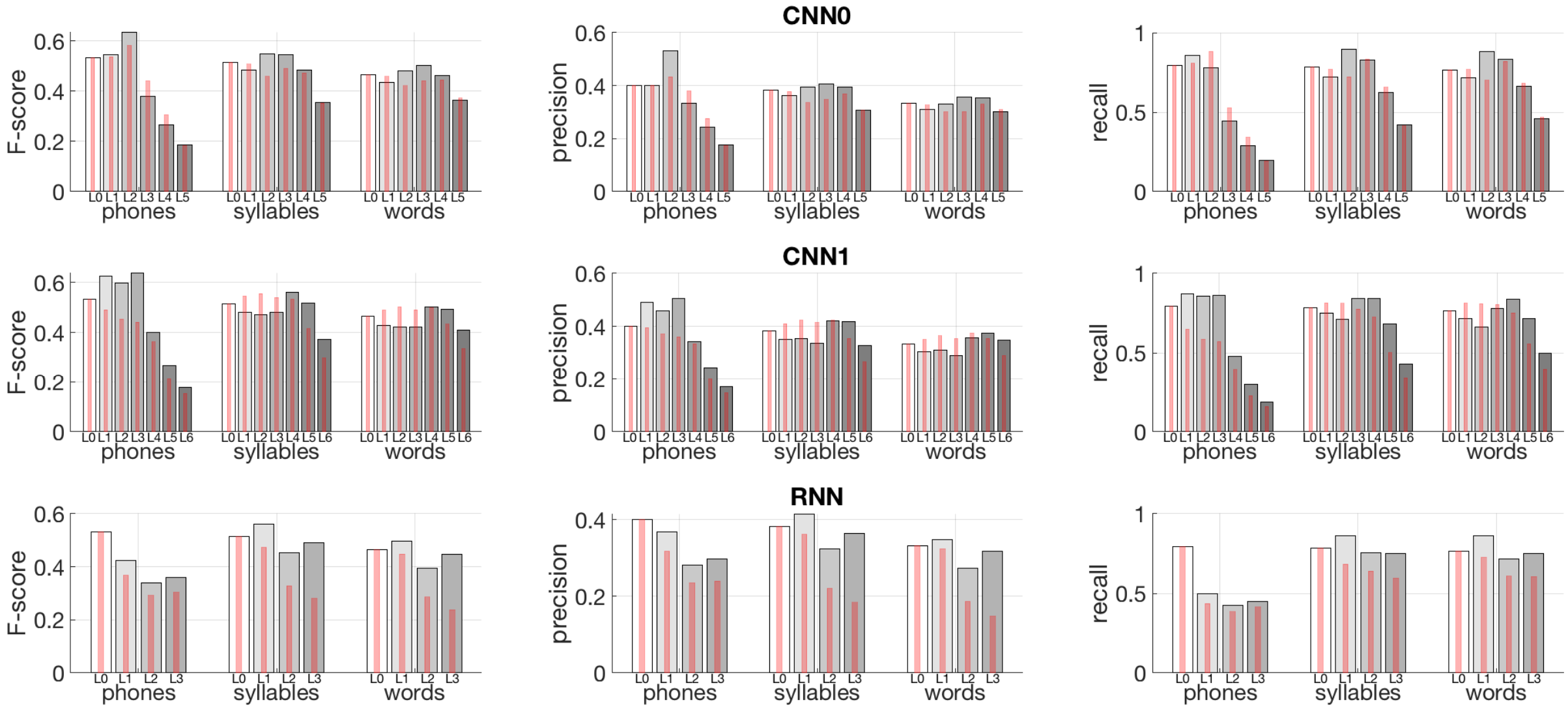

**Figure B.3.** ***Segmentation results for models trained on Places and tested on COCO data when using L2-norm of activation magnitudes as the signal representation.***

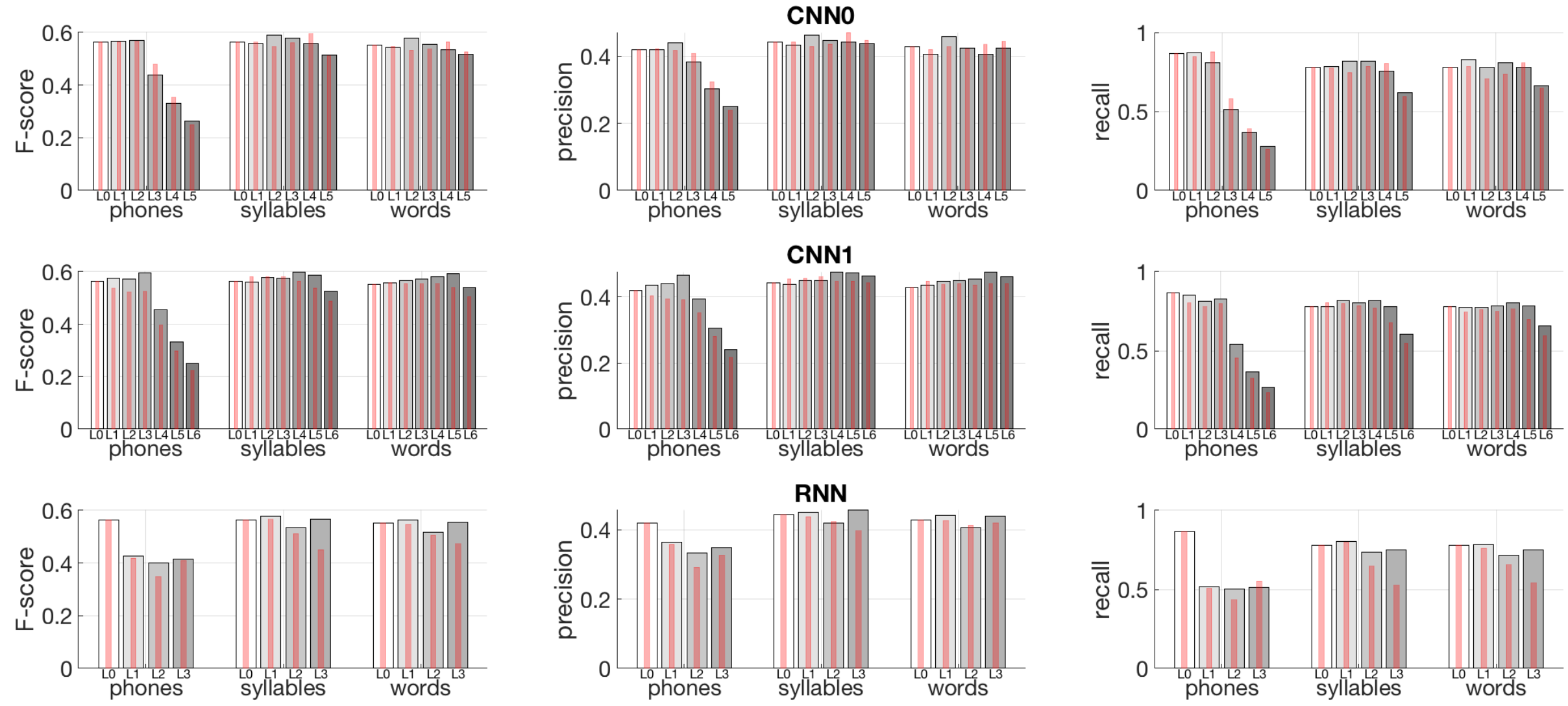

**Figure B.4.** ***Segmentation results for models trained COCO and tested on Brent data when using L2-norm of activation magnitudes as the signal representation.***

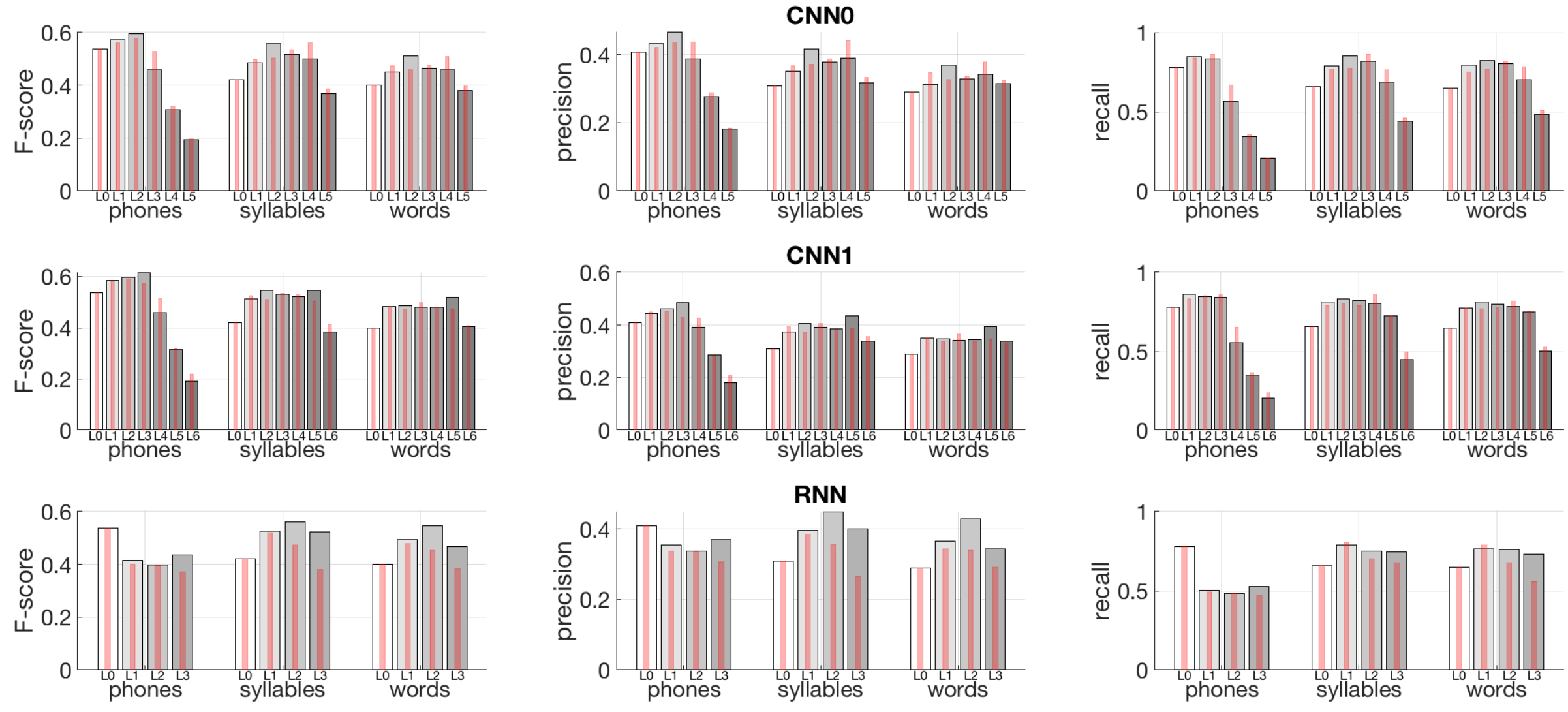

**Figure B.5.** ***Segmentation results from models trained and tested on COCO when using entropy of activation magnitudes as the signal representation. Bars in each subplot represent layer-specific performance metrics for segmenting phones, syllables, and words, respectively. Different layers from input (L0) to last network layer are shown with different shade bars from left to right. Results are shown for F-score (left panels), precision (middle panels), and recall (right panels), and for the three tested models: CNN0 (top), CNN1 (middle), and RNN (bottom). Thin red lines denote chance-level performance with randomized boundary locations.***

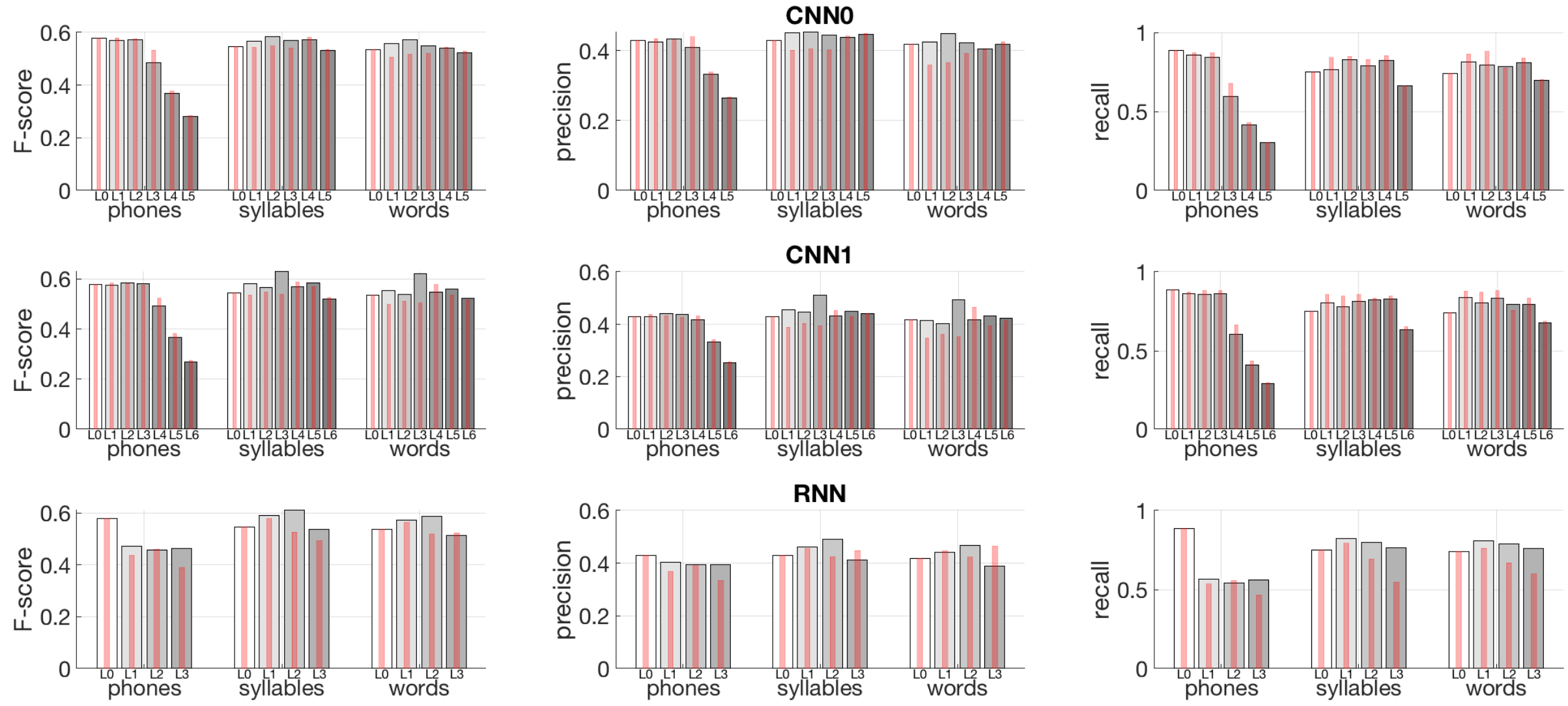

**Figure B.6.** ***Segmentation results from models trained on Places and tested on Brent when using entropy of activation magnitudes as the signal representation.***

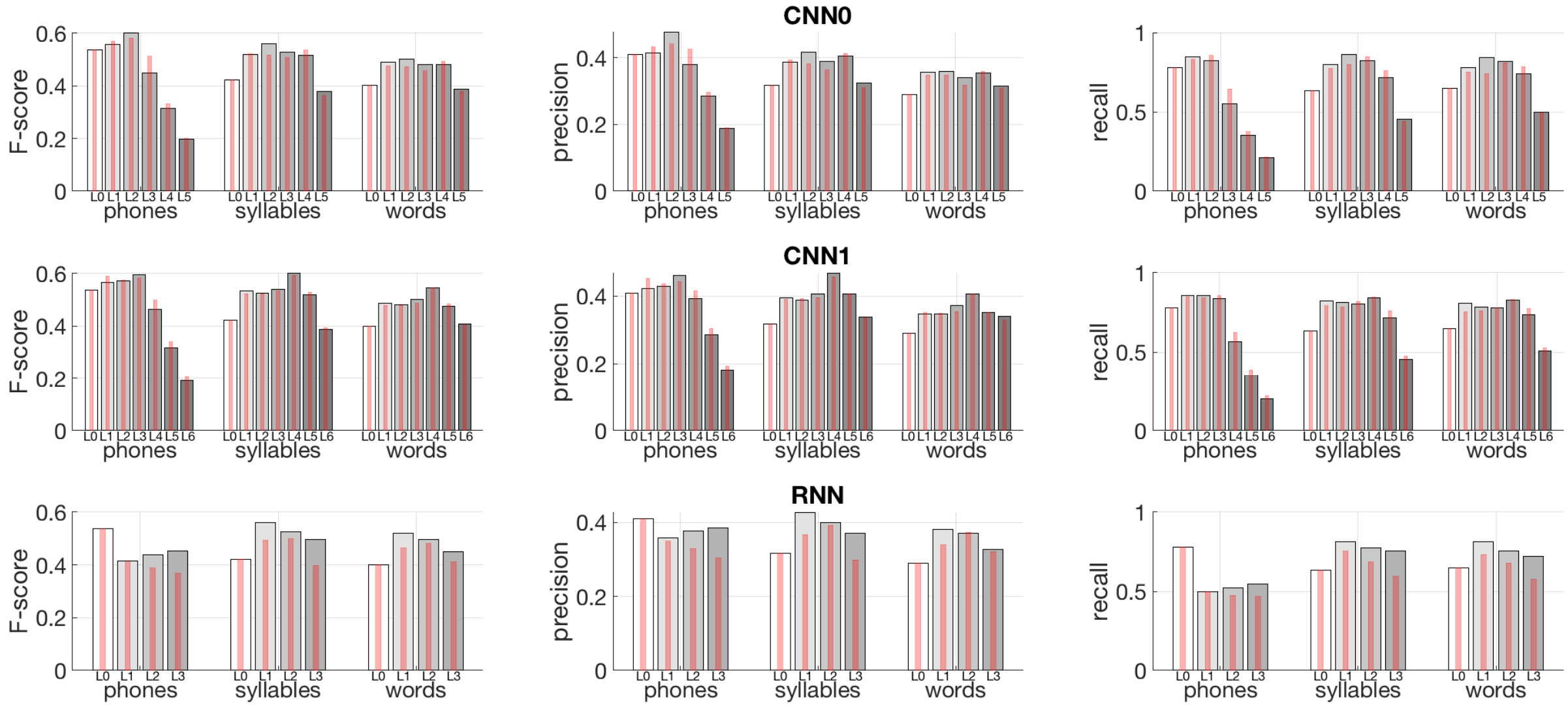

**Figure B.7.** ***Segmentation results from models trained on Places and tested on COCO when using entropy of activation magnitudes as the signal representation.***

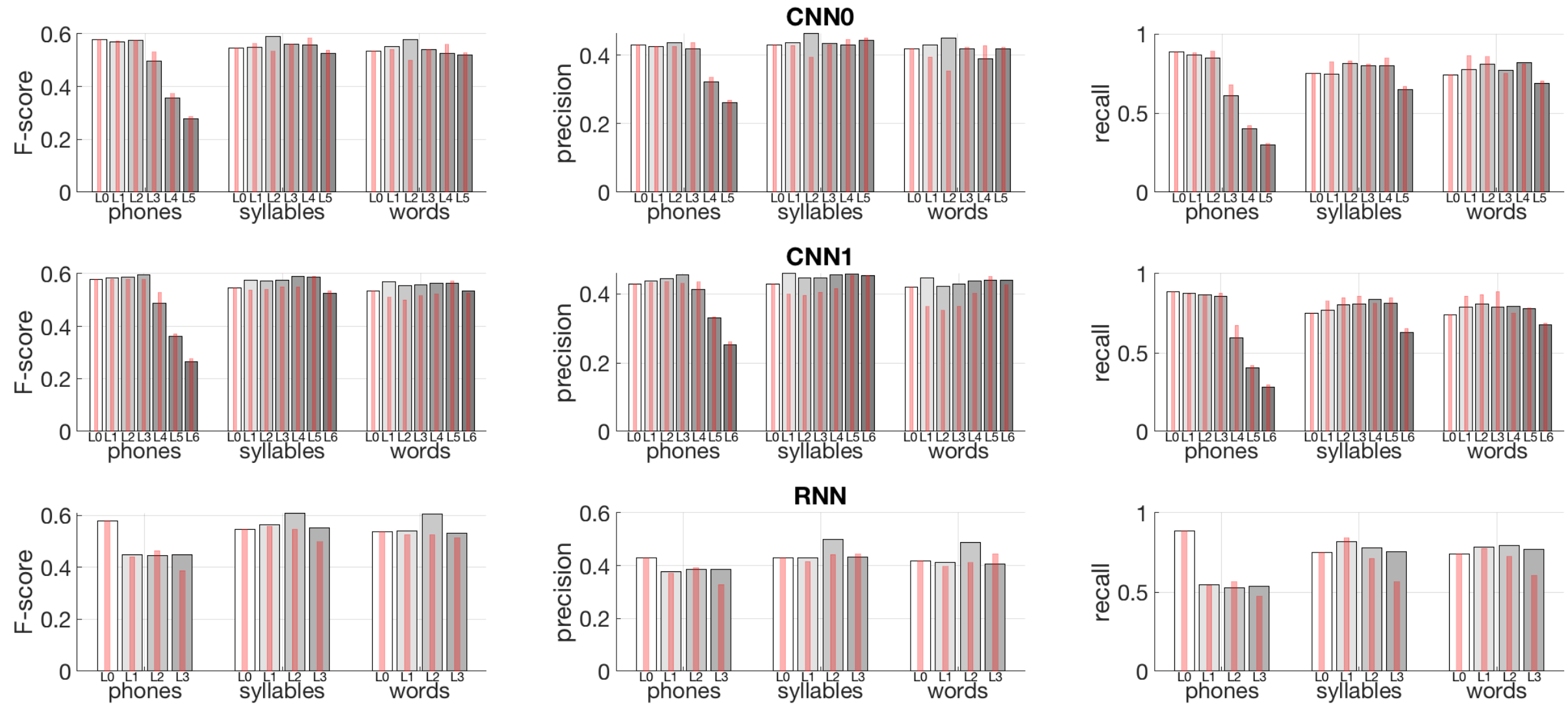

**Figure B.8.** ***Segmentation results from models trained on COCO and tested on Brent when using entropy of activation magnitudes as the signal representation.***

## Appendix C

Wilcoxon rank-sum statistics for SRS and SBERT similarity score distributions from speech-to-speech search are shown in Table C.1.

**Table C.1.** ***Wilcoxon rank-sum statistic for speech-to-speech search for tested models, comparing the distributions consisting of the SRS scores for 5 nearest vs. 5 furthest utterances ("near vs. far") or 5 nearest vs. 5 random utterances ("near vs. random") for each query utterance ($p < 0.001$ for all). Left: SRS results using all content words in the utterances. Middle: SRS results excluding repeating words between query and search result utterances. Right: SBERT results for full captions.***

| | **SRS w. repeating words** | | **SRS wo. repeating words** | | **SBERT** | |
|---|---|---|---|---|---|---|
| **COCO** (df = 24385) | near vs. far | near vs. rand | near vs. far | near vs. rand | near vs. far | near vs. rand |
| **CNN0** | 180.41 | 170.52 | 143.14 | 127.91 | 186.21 | 178.65 |
| **CNN1** | 185.06 | 174.12 | 155.62 | 128.64 | 188.73 | 180.63 |
| **RNN** | 177.65 | 169.28 | 139.16 | 129.79 | 184.11 | 178.26 |
| **Places** (df = 46210) | near vs. far | near vs. rand | near vs. far | near vs. rand | near vs. far | near vs. rand |
| **CNN0** | 192.90 | 169.10 | 146.91 | 119.20 | 209.34 | 184.31 |
| **CNN1** | 208.77 | 182.91 | 151.18 | 119.26 | 226.44 | 198.53 |
| **RNN** | 186.41 | 167.63 | 145.35 | 125.72 | 201.31 | 184.14 |

## License